\documentclass[useAMS,usenatbib,fleqn,twocolumn]{aastex631}
\usepackage{multirow}
\usepackage{graphicx}
\usepackage{xcolor,soul}
\usepackage{amssymb,amsmath}
\usepackage{aas_macros}
\usepackage{tabularx} 
\usepackage[T1]{fontenc} 
\begin{document}

\shorttitle{MAD Spin Evolution}
\title{Multi-messenger Probes of Supermassive Black Hole Spin Evolution}

\shortauthors{Ricarte et al.}
\correspondingauthor{Angelo Ricarte}
\email{angelo.ricarte@cfa.harvard.edu}

\author[0000-0001-5287-0452]{Angelo Ricarte}
\affiliation{Black Hole Initiative at Harvard University, 20 Garden Street, Cambridge, MA 02138, USA}
\affiliation{Center for Astrophysics | Harvard \& Smithsonian, 60 Garden Street, Cambridge, MA 02138, USA}

\author[0000-0002-5554-8896]{Priyamvada Natarajan}
\affiliation{Department of Astronomy, Yale University, 219 Prospect Street, New Haven, CT 06511, USA
}
\affiliation{Black Hole Initiative at Harvard University, 20 Garden Street, Cambridge, MA 02138, USA}

\author[0000-0002-1919-2730]{Ramesh Narayan}
\affiliation{Black Hole Initiative at Harvard University, 20 Garden Street, Cambridge, MA 02138, USA}
\affiliation{Center for Astrophysics | Harvard \& Smithsonian, 60 Garden Street, Cambridge, MA 02138, USA}

\author[0000-0002-7179-3816]{Daniel C. M. Palumbo}
\affiliation{Black Hole Initiative at Harvard University, 20 Garden Street, Cambridge, MA 02138, USA}
\affiliation{Center for Astrophysics | Harvard \& Smithsonian, 60 Garden Street, Cambridge, MA 02138, USA}

\date{\today}

\begin{abstract}

Using the semi-analytic model {\sc Serotina}, we investigate the cosmic spin evolution of supermassive black holes incorporating recent results from general relativistic magnetohydrodynamics simulations of spin-down from relativistic jets. We compare several variations of our model with compiled black hole spin measurements derived from X-ray reflection spectroscopy, correcting for a bias arising from the spin-dependent radiative efficiency of accretion flows. We show that the observed spin distribution is in agreement with a model that includes jet-driven spin-down, a key mechanism that acts to modulate spins across cosmic time at both high and very low specific accretion rates.  The data also clearly prefer models with coherent accretion over models in which accretion disks rapidly switch from prograde to retrograde.  We further predict spin distributions accessible via spatially resolved event horizons by the next-generation Event Horizon Telescope (ngEHT) and Black Hole Explorer (BHEX), as well as gravitational waves by the Laser Interferometer Space Antenna (LISA), each of which offer unique and distinct windows into the population of spinning black holes.  Jet-driven spin-down is most strongly imprinted on the abundance of very highly spinning objects in our model. In addition, we show that the spin distribution sampled by LISA events may contain a signature of the natal spin distribution of heavy seeds, but not of light seeds, offering additional discrimination between these seeding pathways. Spin distributions from these future observed samples can be used to constrain the detailed physical properties of the accretion flow on horizon scales around supermassive black holes. 
\end{abstract}


\section{Introduction}
\label{sec:introduction}

Supermassive black holes (SMBHs) are found at the centers of galaxies, where they are believed to play an important role regulating gas cooling and star formation via radiation, winds, and jets \citep{Kormendy&Ho2013}. All astrophysical black holes (BHs) are believed to be fully described by two parameters, their mass and spin \citep{Kerr1963}. The masses and growth rates of SMBHs across cosmic time are well-constrained from multi-wavelength observations, and have been extensively modeled, but their spin evolution is much more poorly constrained. The spin of a BH encodes its past assembly via accretion and mergers \citep[e.g.,][]{Bardeen1970,Gammie+2004}, determines the radiative efficiency of its accretion disk from which gas is accreted \citep[e.g.,][]{Shapiro&Teukolsky1983}, and can be tapped electromagnetically to power an efficient jet \citep{Blandford&Znajek1977}.

Recent observational and theoretical advances merit a revisiting of the spin evolution of SMBHs in the larger context of their assembly and evolution as a population over cosmic time.  Observationally, the number of SMBHs with spin measurements via X-ray reflection spectroscopy continues to grow \citep[see][for recent reviews]{Reynolds2021,Bambi+2021}. Meanwhile, the Event Horizon Telescope (EHT) has demonstrated the ability to resolve SMBHs accreting at low accretion rates on event horizon scales \citep[e.g.,][]{EHTC+2019a,EHTC+2022a}.  Ongoing development of the EHT through the next-generation EHT (ngEHT) project and Black Hole Explorer (BHEX) are expected to provide spin constraints on a sample of spatially resolved SMBHs in the next decade \citep{Ricarte+2023b}. Finally, the Laser Interferometer Space Antenna (LISA), will provide measurements of the spins of merging BHs from their gravitational wave-forms \citep{Berti+2005,Lang+2006,Lang+2007,Lang+2008,Lang+2011,Baibhav+2020,Bhagwat+2022,Watarai+2024}.

Seminal SMBH spin studies considered their growth from thin disk accretion and BH-BH mergers, predicting a population of mostly highly-spinning SMBHs if accretion is coherent, or low-spins otherwise \citep{Volonteri+2005,King+2008,Volonteri+2013}. In addition, BH spin is implicated in the formation of jets and their radio properties. Spin evolution models are of interest for testing the spin paradigm of radio-loudness, wherein a minority of spinning SMBHs are hypothesized to produce the observed radio-loud population \citep{Moderski+1998,Tchekhovskoy+2010,Garofalo+2010}. Several authors have predicted spin signatures in galaxy morphology, finding that the larger contribution of BH-BH mergers in the mass assembly histories of more massive galaxies can drive the spin distribution in elliptical galaxies \citep{Volonteri+2007,Barausse2012,Sesana+2014,Izquierdo-Villalba+2020}. SMBH spin can now be self-consistently tracked in some galaxy- or cosmological-scale simulations, oftentimes directly linked to the feedback efficiency adopted for the actively accreting SMBH \citep{Dubois+2014,Fiacconi+2018,Bustamante&Springel2019,Beckmann+2019,Dubois+2021,Talbot+2021,Massonneau+2023b,Dong-Paez+2023,Koudmani+2024,Sala+2024,Beckmann+2024b} Although such simulations cannot resolve the turbulent ISM on sub-pc scales, they typically produce highly spinning AGN in line with expectations from coherent accretion.

Additional insights from data into the spins of SMBHs have been provided by recent horizon-scale constraints on accretion from the EHT data. The compatibility of EHT observations with accretion at extremely low rates have spurred deeper theoretical exploration of the accretion process via ``Magnetically Arrested Disks'' (MADs), which are currently the preferred models for data of both M87 and Sgr A$^*$ \citep{EHTC+2019e,EHT_SgrA_V,EHTC+2022e,EHTC+2024c}. These MAD models are characterized by magnetic fields strong enough to dominate the dynamics of the accretion flow \citep{Bisnovatyi-Kogan&Ruzmaikin1974,Igumenshchev+2003,Narayan+2003}, in contrast with their weakly magnetized ``Standard and Normal Evolution'' (SANE) counterparts \citep{Narayan+2012,Sadowski+2013}.  MAD GRMHD simulations predict jets powered by the \citet{Blandford&Znajek1977} mechanism that can be so efficient that the equilibrium spin is approximately 0 \citep{McKinney+2012,Narayan+2022,Lowell+2024}.  \citet{Ricarte+2023d} found that the BZ mechanism can also power jets and spin down BHs with accreting at Eddington ratios $\gtrsim 0.3$, arriving at a ``sub-grid'' model for spin evolution that is applicable across Eddington ratios.

In light of these recent observational and theoretical advances, in this work we implement the derived novel spin evolution formulae into {\sc Serotina}, a semi-analytic model (SAM) for SMBH evolution that tracks a population of SMBHs from the seeding epoch to the present day \citep{Ricarte&Natarajan2018a,Ricarte&Natarajan2018b,Ricarte+2019b}. These GRMHD simulation derived formulae of \citet{Ricarte+2023d} self-consistently predict spin evolution and jet power as a function of accretion rate. By coupling spin evolution and jet power in a cosmological simulation, \citet{Beckmann+2024b} recently found that jet-driven spin-down at low accretion rates contributes to increased scatter in the SMBH population.  In this work, we also include spin-down at high accretion rates, using a flexible semi-analytic modeling methodology that allows us to test several model variations.  By forward-modeling different selection effects, we connect our findings to multi-messenger probes of SMBH spin including X-ray reflection spectroscopy spin measurements, spins accessible to spatially resolved imaging by the EHT and its extensions, and the spins of SMBH merger events observable by LISA. We demonstrate that spin distributions offer a powerful and more robust discriminant of models for the formation and evolution of SMBHs compared to other observational probes.

The outline of this paper is as follows.  First, in \autoref{sec:methodology}, the ingredients and features of {\sc Serotina}, our model for tracking cosmological SMBH evolution including spins is described. We then present our results \autoref{sec:results}, which includes a comparison with X-ray reflection spectroscopy derived spins (\autoref{sec:x-rays}); predictions for spins that can be probed by extensions to the EHT (\autoref{sec:ngEHT}); and LISA predictions (\autoref{sec:lisa}), each of which we find are sensitive to surprisingly disjoint parts of the population. We then discuss predictions for jet-mode feedback across cosmic time and how our parameters map to observables in  \autoref{sec:discussion}. Our key results are summarized in \autoref{sec:conclusions}.

\section{Serotina and Spin Evolution}
\label{sec:methodology}

First developed in \citet{Ricarte&Natarajan2018a}, {\sc Serotina} is the new name for our augmented semi-analytic model for SMBH-galaxy-halo co-evolution.\footnote{{\it Prunus Serotina} is the black cherry tree, which bears resemblance to a high-resolution merger tree with massive seeds.}  This model focuses on the growth history of SMBHs alone, skipping to the galaxy properties for a given dark matter halo using empirically determined relationships. The only galaxy properties used in this model are the stellar mass and effective radius, from which the velocity dispersion $\sigma$ is estimated \citep[see Section 3.3 of][]{Ricarte&Natarajan2018a}.  It is important to calculate $\sigma$ of the stellar component of the galaxy distinctly from $\sigma$ of the dark matter halo, especially in the most massive halos in order to reproduce reasonable values of $\sigma$ in cluster environments and to match the observed AGN downsizing. The model assumes that the $M_\bullet-\sigma$ relation is the fundamental astrophysical relationship linking SMBHs to their host galaxies. Compared to our most recent iteration of this model presented \citep{Ricarte+2019b}, the new ingredients included in {\sc Serotina} are the following:

\begin{itemize}
    \item {\sc Serotina} now tracks spin using fitting formulae derived from recent GRMHD simulations, as described in \autoref{sec:spin_recipes}.
    \item Because spin evolution depends in detail on how the accretion rate evolves with time, we have implemented a power-law decline in the accretion rate \citep[similar to][] {Hopkins+2006b,Volonteri+2013}.  This replaces the ``steady'' mode we adopted in previous iterations of the model.
    \item A SMBH's radiative efficiency now depends on both spin and the Eddington ratio, based on analytic models.
    \item Since we have found that mergers dominate SMBH growth in this model for the most massive SMBHs, we now explicitly also account for the mass lost during mergers to gravitational waves.
\end{itemize}

\subsection{Merger Trees}

Dark matter merger trees form the backbone of the model as before. These are calculated using the analytic \citet{Press&Schechter1974} binary tree formalism, as implemented by \citet{Parkinson+2008} following calibration to the Millenium dark matter only N-body simulations \citep{Springel+2005a}. Compared to merger trees taken directly from N-body simulations, Press-Schechter merger trees have the disadvantage of lacking direct spatial information. On the other hand, they can be generated much more rapidly, achieve high resolution by extending to arbitrarily small halo masses, and avoid issues associated with halo finding in an N-body simulation. As in our previous work, we sample 23 halo masses between $10.6 \leq \log M_h \; [M_\odot] \leq 15$, each simulated 20 times to account for cosmic variance. The minimum halo mass in the merger tree is redshift and host halo mass dependent, reaching a host halo mass of $5 \times 10^6 \ M_\odot$ at $z=20$ \citep[see][]{Ricarte&Natarajan2018a}.

\subsection{SMBH Seeding}

The seeding of SMBHs is active field of research with many outstanding questions including the details of the how the first BHs form, how seeds migrate to the centers of galaxies, to how they can grow efficiently enough to explain the massive populations we observe at $z \sim 7$ \citep[see][for recent reviews]{Natarajan+2019,Woods+2019,Inayoshi+2020}. We adopt simple seeding prescriptions to apply in the metal-poor universe between redshifts $15 \leq z \leq 20$.  Low-mass and abundant ``light'' seeds can originate from the remnants of Pop III star formation, and we preserve the power-law IMF adopted in \citep[][]{Ricarte&Natarajan2018b}, with $dn/d \log M_\bullet \propto M_\bullet^{-0.3}$ between 30 and 100 solar masses in halos exceeding a 3.5$\sigma$ peak in the dark matter distribution \citep[][]{Hirano+2014,Stacy+2016}.  Alternatively, we assume that intermediate-mass and rarer ``heavy'' seeds from direct collapse can form according to the model of \citet{Lodato&Natarajan2006,Lodato&Natarajan2007}, based on global properties of proto-galactic disks and their instabilities. The claims in our work here regarding discrimination between light and heavy seeds arise from broad differences in mass and abundance, irrespective of finer details: light seeds are more abundant and have lower masses, while heavy seeds are rarer and have higher masses.

Furthermore, we assume that seeds are always initialized with $a_\bullet=0$. This choice is justified under the assumption that seeds form in a highly super-Eddington manner, and the equilibrium spin of highly super-Eddington accretion disks is expected to be close to 0 \citep{Ricarte+2023d,Lowell+2024}.  

\subsection{BH Mergers}

BH mergers have been the explored extensively in both cosmological and idealized simulations, however their implementation is complicated by our poor understanding of the characteristic timescales involved in the merger of the corresponding dark matter halos and host galaxies prior to the merger of the central BHs. The time elapsed between a galaxy merger and the merger of the central BHs hosted in them varies with environment and is currently a rich and very active area of research. Significant delays in this sequence of mergers are possible and poorly constrained observationally on both galactic \citep[e.g.,][]{Izquierdo-Villalba+2020,Ricarte+2021,DiMatteo+2023} and sub-kpc scales \citep[e.g.,][]{Milosavljevic&Merritt2001,Dosopoulou&Antonini2017}.  Here we, simply assume that the SMBHs merge promptly with probability $p_\mathrm{merge} \in [0,1]$ following the merger of their host halos, which occurs after a dynamical friction timescale, calculated as a function of mass ratio via \citet{Boylan-Kolchin+2008}.  In previous work where the same merger probability prescription was adopted, we discussed how $p_\mathrm{merge}=0.1$ helped the most massive SMBHs from overgrowing past the local $M_\bullet-\sigma$ relation via SMBH-SMBH mergers \citet{Ricarte&Natarajan2018a}. Although a direct comparison with our model predictions is beyond the scope of this current paper, the large amplitude of the nHz gravitational wave background reported by Pulsar Timing Array experiments like NANOGrav prefers models with efficient merging, at least in local massive halos \citep{Izquierdo-Villalba+2022,NANOGrav+2023}.  

When a merger occurs, we compute the spin of the remnant according to the fitting formulae of \citet{Rezzolla+2008}.  We assume that the secondary BH merges at a random orientation with respect to both its own and the primary's angular momentum axis.  We account for mass loss via gravitational waves, computing the mass of the remnant according to the fitting formulae of \citet{Lousto+2010}, which are based on numerical general relativistic calculations \citep{Lousto+2014}.  Lacking a framework for orbital evolution of off-nuclear SMBHs, we neglect the gravitational wave recoil from asymmetric binary coalescences.

\subsection{BH Growth and Accretion}

The formulae adopted in this section originate from a variety of sources and require a heterogeneous set of accretion rate definitions.  We denote the mass inflow rate through the disk as $\dot{M}_d$, which is distinct from the rate of change of SMBH rest mass, $\dot{M}_\bullet$.  We define the standard Eddington ratio:

\begin{equation}
    f_\mathrm{Edd} \equiv \frac{L_\mathrm{bol}}{L_\mathrm{Edd}},
    \label{eqn:f_Edd}
\end{equation}

\noindent where $L_\mathrm{bol}$ is the bolometric luminosity and $L_\mathrm{Edd}$ is the Eddington luminosity, 

\begin{equation}
    L_\mathrm{Edd} = \frac{4 \pi M_\bullet m_p c}{\sigma_T},
    \label{eqn:L_Edd}
\end{equation}

\noindent where $m_p$ is the proton mass, $c$ is the speed of light, and $\sigma_T$ is the Thomson cross-section. The fitting functions of \citet{Ricarte+2023d} are formulated in terms of a normalized accretion rate that we denote as:

\begin{equation}
    r_\mathrm{Edd} \equiv \frac{\dot{M}_d}{\dot{M}_\mathrm{Edd,thin}},
    \label{eqn:r_Edd}
\end{equation}

\noindent where $\dot{M}_\mathrm{Edd,thin}$ is the mass accretion rate that would produce the Eddington luminosity, if the radiative efficiency were given by that of a \citet{Novikov&Thorne1973} thin disk.  

If one were to fix $M_\bullet$ and either $f_\mathrm{Edd}$ or $r_\mathrm{Edd}$, $\dot{M}_d$ would be dependent on the SMBH spin through the radiative efficiency. This is inconvenient in cases where galactic inflows govern the inflow rate or when the radiative efficiency should itself be calculated as a function of accretion rate.  Thus, we also define a mass-normalized specific accretion rate 

\begin{equation}
    \dot{m} = \frac{0.1\dot{M}_dc^2}{L_\mathrm{Edd}},
    \label{eqn:m_Edd}
\end{equation}

\noindent which is the mass inflow rate compared to the value that would produce the Eddington luminosity assuming a fiducial radiative efficiency of 0.1.

There are two efficiencies for our accretors, the radiative efficiency that relates luminosity to inflow rate, 

\begin{equation}
    \epsilon = \frac{L_\mathrm{bol}}{\dot{M}_d c^2},
    \label{eqn:epsilon}
\end{equation}

\noindent and the jet efficiency 

\begin{equation}
    \eta = \frac{P_\mathrm{jet}}{\dot{M}_d c^2},
\end{equation}

\noindent where $P_\mathrm{jet}$ is the jet power.  Combining results and taking into account the loss of rest mass energy from the system, the SMBH mass evolves via:

\begin{equation}
    \dot{M}_\bullet = (1-\epsilon-\eta)\dot{M}_d = \frac{1-\epsilon-\eta}{\epsilon}\frac{f_\mathrm{Edd}L_\mathrm{Edd}}{c^2}. \label{eqn:mass_evolution}
\end{equation}

Interestingly, there is no requirement that $1-\epsilon-\eta > 0$.  That is, for systems with efficient enough jets, it is known that the \citet{Blandford&Znajek1977} mechanism can reduce the rest mass energy of a SMBH (down to a finite ``irreducible mass'').  In practice, we find this is a very small effect that only occurs for brief periods.

\subsubsection{Accretion Rates}

We have experimented with several accretion/growth prescriptions in \citet{Ricarte&Natarajan2018a} and \citet{Ricarte+2019b}, finding the best match to determined luminosity functions drawing Eddington ratios from observed distributions for broad line quasars \citep{Tucci&Volonteri2017} during major mergers, and no accretion otherwise.  However, more direct comparisons to growing SMBHs in dwarf galaxy populations have revealed a need to maintain non-negligible accretion at some level outside of efficient merger driven growth \citep{Chadayammuri+2023,Sacchi+2024}. In this paper, SMBH Eddington ratios are assigned in the following manner, similar to \citep{Volonteri+2013}:

\begin{itemize}
    \item All seeds are initialized with $\dot{m}=0$.
    \item When a halo merger of mass ratio $\geq 0.1$ occurs (defined as our criterion for a major merger), SMBHs are placed into the ``burst'' mode.  While in the burst mode, values of $\dot{m}$ are drawn at each time step from the log-normal fits of \citet{Tucci&Volonteri2017} for Type I AGN.\footnote{Observationally, it is $f_\mathrm{Edd}$ that is constrained rather than $\dot{m}$.  However, we choose to assign $\dot{m}$ values instead of $f_\mathrm{Edd}$ values because the function $f_\mathrm{Edd}(a_\bullet,\dot{m})$ is not invertible due to the discontinuity at $\dot{m}_\mathrm{crit}$ (\autoref{fig:recipes}, second panel).}  SMBHs exit the burst mode and enter the ``decline'' mode once they $M_\mathrm{\bullet,\mathrm{max}}(\sigma)$, defined below. 
    \item While in the decline mode, we set $\dot{m} = (1+(t/10^7 \ \mathrm{yr})^{2})^{-1}$ \citep{Hopkins+2006b,Hopkins+2006}.\footnote{Note that this temporal power law decline is distinct from the power law assumed for a universal ``steady mode'' Eddington ratio distribution in \citet{Ricarte&Natarajan2018a,Ricarte&Natarajan2018b}.}  To add a realistic amount of scatter, we assume that $\dot{m}$ at each $t$ is log-normally distributed using the same log-normal width as for Type I AGN at the same epoch from \citet{Tucci&Volonteri2017}. 
\end{itemize}

In \citet{Ricarte&Natarajan2018a}, we tuned two parameters in $M_\mathrm{\bullet,\mathrm{max}}(\sigma)$ by hand to reproduce the observed $M_\bullet-\sigma$ relation and luminosity functions across cosmic time.  This is given by 

\begin{equation}
    \log M_{\bullet,\mathrm{max}}(\sigma) = 8.45 + 5.0 \left( \frac{\sigma}{200 \; \mathrm{km}\mathrm{s}^{-1}} \right) \label{eqn:M_max}
\end{equation}

\noindent These are the only parameters tuned to produce reasonable match with observables.

\subsubsection{Radiative and Jet Efficiencies}

In our model, we first specify $\dot{m}$.  Then, depending on $\dot{m}$, the accretion flow falls into one of the following three different accretion regimes

\begin{itemize}
    \item Hot Accretion:  Radiatively inefficient and kinetically efficient accretion occurs for $\dot{m} < \dot{m}_\mathrm{crit}$, where $\dot{m}_\mathrm{crit} \sim 10^{(-2)-(-3)}$ is a free parameter.
    \item Radiatively Efficient:  We model $\dot{m}_\mathrm{crit} < \dot{m} \lesssim 0.3$ as a standard radiatively efficient and kinetically inefficient thin disk with no jet.
    \item Super-Eddington:  For $\dot{m} \gtrsim 0.3$, our fitting formulae provide a continuous transition towards a more radiatively inefficient and kinetically efficient disk.
\end{itemize}

The standard \citet{Novikov&Thorne1973} thin disk has a radiative efficiency $\epsilon = \epsilon_\mathrm{NT}$ given by

\begin{equation}
    \epsilon_\mathrm{NT}(a_\bullet) = 1 - \left[1-\frac{2}{3r_\mathrm{ISCO}(a_\bullet)} \right]^{1/2}.
    \label{eqn:eps_nt}
\end{equation}

\noindent Here, $r_\mathrm{ISCO}$ is the spin-dependent radius of the Innermost Stable Circular Orbit (ISCO) in geometrized units \citep[see e.g.,][]{Bardeen1970}.  Since this model lacks a jet, $\eta = 0$.

We modify this radiative efficiency depending on $\dot{m}$, taking into account losses in efficiency at both very high and very low accretion rates.  In general, the radiative efficiency in our model is given by:

\begin{equation}
    \epsilon(a_\bullet, \dot{m}) = \epsilon_\mathrm{NT}(a_\bullet) \cdot \begin{cases} 
    \frac{1}{100}, & \dot{m} < \dot{m}_\mathrm{crit} \\
    1, & \dot{m}_\mathrm{crit} \leq \dot{m} < 5 \\
    \frac{5}{\dot{m}} \left[ 1+ \ln \left( \frac{\dot{m}}{5} \right) \right], & \dot{m} \geq 5
    \end{cases}
    \label{eqn:radiative_efficiency}
\end{equation}

For $\dot{m} < \dot{m}_\mathrm{crit}$, the factor of 100 is selected to match the radiative efficiency of Sgr A$^*$, with $\dot{m}_\mathrm{Edd} \sim 10^{-7}$ and $\epsilon \sim 0.003$ \citep{EHT_SgrA_V}.  Meanwhile, we continuously transition from the radiatively efficient \citet{Novikov&Thorne1973} thin disk at $\dot{m}=1$ to the super-Eddington behavior of analytic slim disk models \citep{Mineshige+2000,Watarai+2001}.

The jet efficiency $\eta$ depends not only on $a_\bullet$ and $\dot{m}$, but also the magnetic field state.  For SANE models, we set $\eta=0$.\footnote{GRMHD models in the literature are generally described as SANE as long as they do not reach horizon magnetic fluxes as high as those of MAD models, but they can still power less efficient jets \citep[e.g.,][]{EHT5}.  However, such jets can be made arbitrarily weak by decreasing the magnetic field strength. For the purposes of this work, SANE models do not produce jets.}  For MAD models, we adopt the fitting formulae of \citet{Ricarte+2023d}, which includes BZ jets at both $\dot{m} < \dot{m}_\mathrm{crit}$ and $\dot{m} \gtrsim 0.3$.  Using GRMHD simulations \citet{Tchekhovskoy+2010} derive the following formula for jet efficiency for a given value of $a_\bullet$:

\begin{equation}
    \eta_\mathrm{BZ} = \frac{\kappa}{4\pi}\phi^2 \Omega_H^2 (1+1.38\Omega_H^2-9.2\Omega_H^4), \label{eqn:eta_bz}
\end{equation}

\noindent where $\kappa$ is a constant set to 0.05, and $\Omega_H=|a_\bullet|/(2(1+\sqrt{1-a_\bullet^2}))$ is the angular velocity of the event horizon.  Using ideal GRMHD simulations in the hot accretion regime, \citet{Narayan+2022} obtained a fitting function for $\phi$ in the MAD state that we apply here:

\begin{equation}
    \phi_\mathrm{hot} = -20.2 a_\bullet^3 - 14.9 a_\bullet^2 + 34 a_\bullet + 52.6. \label{eqn:phi_hot}
\end{equation}

In the super-Eddington regime, \citet{Ricarte+2023d} found that $\phi$ could be estimated as:

\begin{equation}
    \phi = \phi_\mathrm{hot} \frac{\xi}{1+\xi}, \label{eqn:phi_superEdd}
\end{equation}

\noindent where $\xi = (r_\mathrm{Edd}/1.88)^{1.29}$.  \autoref{eqn:phi_superEdd} approaches 0 as $r_\mathrm{Edd} \to 0$ and approaches $\phi_\mathrm{hot}$ as $r_\mathrm{Edd} \to \infty$.  That is, it transitions smoothly to a thin disk at low accretion rates and approaches the ideal GRMHD result at high accretion rates.

\subsection{Spin Evolution by Accretion and Jets}
\label{sec:spin_recipes}

We parameterize spin evolution using the spin-up parameter $s$ from \citet{Gammie+2004},

\begin{equation}
    s \equiv \frac{d a_\bullet}{dt}\frac{M_\bullet}{\dot{M}_d} \label{eqn:s}
\end{equation}

\noindent which describes the relative evolutionary build-up of mass and spin.  We adopt different equations for spin evolution depending on whether accretion disks are assumed to be MAD or SANE. Regardless of the type of accretion flow assumed, the spin parameter is capped at 0.998, the maximum value achievable by a thin disk due to a counteracting torque from its own radiation \citep{Thorne1974}.

\subsubsection{SANE (Standard Thin Disk)}

For SANE accretion flows, we assume that the standard thin disk accretion equations apply.  Following \citet{Shapiro2005} for a thin disk that terminates at the innermost stable circular orbit (ISCO),

\begin{equation}
    s_\mathrm{SANE}(a_\bullet) = l - 2 a_\bullet e \label{eqn:s_sane}
\end{equation}

\noindent where $l$ and $e$ represent the specific angular momentum and energy of particles at the ISCO, which is spin-dependent.

\subsubsection{MAD (Strong Magnetic Fields)}

For MAD accretion flows, we evolve the spin according to the formulae of \citet{Ricarte+2023d}, where changes to the disk dynamics and angular momentum loss due to the BZ mechanism are taken into account.  Building upon \citet{Moderski&Sikora1996} and \citet{Lowell+2024}, these formulae split the spinup parameter into its hydrodynamical and electromagnetic components,

\begin{equation}
    s_\mathrm{MAD} = s_\mathrm{HD,MAD} + s_\mathrm{EM,MAD}. \label{eqn:s_mad}
\end{equation}

\citet{Ricarte+2023d} modeled the changes to the accretion flow as a function of Eddington ratio using radiative GRMHD simulations with $0.3 \leq r_\mathrm{Edd} \leq 40$.  They found that the spinup in these simulations could be characterized by:

\begin{equation}
    s_\mathrm{HD,MAD} = \frac{s_\mathrm{SANE} + s_\mathrm{min} \xi}{1+\xi}, 
    \label{eqn:sHD}
\end{equation}

\noindent where $\xi = 0.017 \; r_\mathrm{Edd}$ and $s_\mathrm{min} = 0.86 - 1.94a_\bullet$, and 

\begin{align}
    s_\mathrm{EM,MAD} = -\mathrm{sign}(a_\bullet)\,\eta_\mathrm{BZ} \left( \frac{1}{k \Omega_H} - 2 a_\bullet \right).
\end{align}

As the Eddington ratio increases, the magnetic flux confined onto the BH horizon increases.  The inflow grows more radial and the BZ jet becomes more efficient, both of which contribute to altering $s$.  As the Eddington ratio decreases, these formulae are designed to asymptote towards the Novikov-Thorne thin disk solution.  Consequently, $s_\mathrm{MAD} \approx s_\mathrm{SANE}$ for $r_\mathrm{Edd} \lesssim 0.3$.   

For $\dot{m} < \dot{m}_\mathrm{crit}$, we switch to a polynomial fit originating from ideal GRMHD simulations appropriate for hot accretion \citep{Narayan+2022}, given by

\begin{equation}
\begin{aligned}
    s_\mathrm{hot}(a_\bullet) = &0.45 - 12.53 a_\bullet -7.80 a_\bullet^2 + 9.44 a_\bullet^3 \\
    &+5.71 a_\bullet^4 - 4.03 a_\bullet^5. \label{eqn:s_hot}
\end{aligned}
\end{equation}

\noindent As $f_\mathrm{Edd} \to \infty$, $s_\mathrm{MAD}$ resembles $s_\mathrm{hot}$. 

\begin{figure*}
    \centering
    \includegraphics[width=\textwidth]{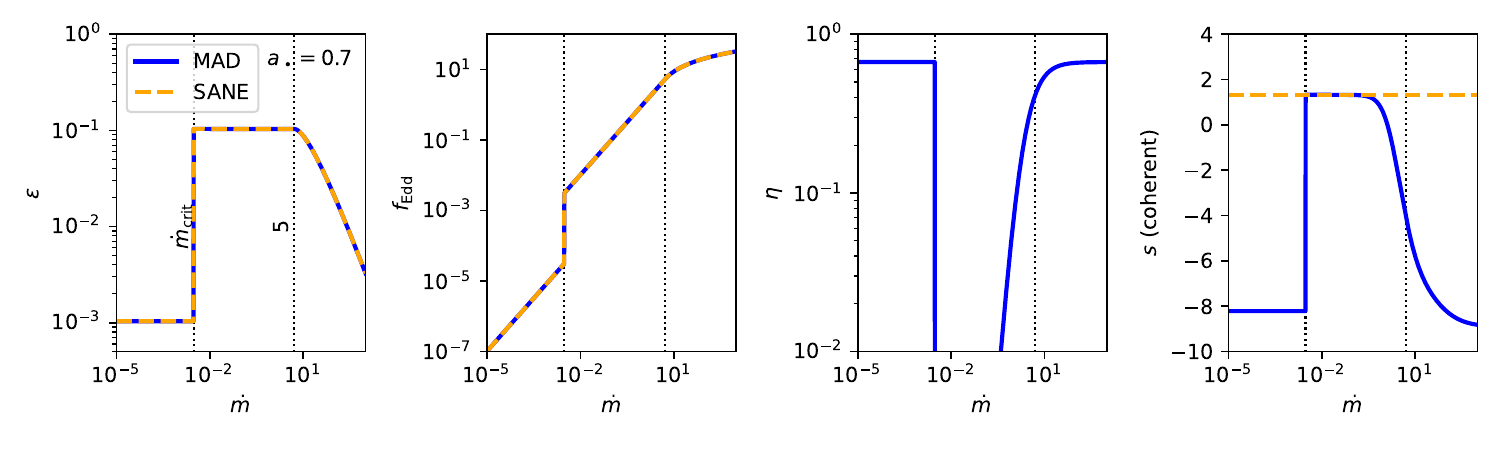}
    \caption{Dependence of selected key quantities (radiative efficiency $\epsilon$, observed Eddington ratio $f_\mathrm{Edd}$, jet efficiency $\eta$, and spin-up parameter $s$) on specific accretion rate ($\dot{m}$).  Our MAD formulae are shown in blue, while our SANE formulae are shown in orange, for a fiducial spin value of $a_\bullet=0.7$. An abrupt transition occurs at $\dot{m}_\mathrm{crit}$, which is set to 0.003 in this illustration.  Both our MAD and SANE formulae account for drops in radiative efficiency at the $\dot{m}<\dot{m}_\mathrm{crit}$ and $\dot{m} > 5$, but only our MAD formulae account for strong magnetic fields that can power BZ jets and modify the dynamics of the inflow.  For $\dot{m}_\mathrm{crit} \leq \dot{m} \lesssim 0.3$, the MAD and SANE models are indistinguishable.}
    \label{fig:recipes}
\end{figure*}

The results of this section so far are summarized in \autoref{fig:recipes}.  The radiative efficiency ($\epsilon$), observed Eddington ratio ($f_\mathrm{Edd}$), jet efficiency ($\eta$), and spin-up parameter ($s$) are each plotted as a function of specific accretion rate ($\dot{m}$).  Our MAD formulae are plotted in blue and our SANE formulae are plotted in orange, fixing $a_\bullet=0.7$ and $\dot{m}_\mathrm{crit}=0.003$.  The radiative efficiency $\epsilon$ and therefore the Eddington ratio $f_\mathrm{Edd}$ both decrease at $\dot{m}<\dot{m}_\mathrm{crit}$ and $\dot{m} > 5$, and do not depend on the magnetic field state.  Our MAD models account for changes to the disk dynamics due to strong magnetic fields that power a BZ jet, but our SANE models do not.  Between $\dot{m}_\mathrm{crit} \leq \dot{m} \lesssim 0.3$, the MAD and SANE models are indistinguishable.

\subsubsection{Prograde or Retrograde?}

The formulae in the preceding subsections support both prograde and retrograde accretion disks with respect to the angular momentum of the SMBH.  Throughout this work, $a_\bullet < 0$ denotes a retrograde accretion disk.  In our model, there are two free parameters that govern the incidence of retrograde accretion disks.  

We implement a ``disruptive'' tag that can flip SMBH disks retrograde during mergers.  When a SMBH merger of mass ratio 1:10 or larger occurs, and it coalesces with an orbit retrograde with respect to the primary spin, we flip the sign of $a_\bullet$ in models with ``disruptive'' instead of ``gentle'' mergers.  In this picture, the major merger may cause a reorientation of the angular momentum of the gas fueling the SMBH.  

We implement a ``coherent'' tag to describe whether the accretion disk maintains the same angular momentum direction over cosmic time.  If accretion is ``incoherent,'' we average the spinup parameters for the prograde and retrograde cases:  $s=(s(a_\bullet)-s(-a_\bullet))/2$, where our conventions require the insertion of a minus sign.  Since we initialize BHs with a spin of 0, if a model is ``incoherent,'' $|a_\bullet|>0$ occurs if and only if a BH-BH merger has occurred.  When a snapshot is saved for a model with incoherent accretion, the sign of $a_\bullet$ is chosen randomly.  As mentioned in \autoref{sec:introduction}, cosmological simulations that track spin find little evidence of incoherent accretion.  On the other hand, megamaser galaxies with jets show significant alignment between the jet and the megamaser disk on pc-scales, but not with larger scale disk structures \citep{Greene+2013,Kamali+2019,Dotti+2024}.  Our own galactic center exhibits a complex angular momentum structure \citep{Murchikova+2019}, and disks can tear and precess due to the Bardeen-Petterson effect \citep{Nixon+2012,Liska+2021}.  The flexibility of our approach allows us to consider both coherent and incoherent models.

Lacking appropriate formulae, we do not consider tilted accretion disks in this work.  This assumption is well-justified for thin disks at high Eddington ratios, where Bardeen-Petterson alignment acts on a timescale shorter than the Salpeter timescale \citep{Natarajan&Pringle1998}.  For geometrically thick disks, the magnetic field may align the disk in some systems \citep{Chatterjee+2023}, but not always.  We expect the magnetic flux of significantly tilted disks to be suppressed compared to their perfectly aligned or anti-aligned counterparts, which would reduce the jet efficiency.  

\begin{figure*}
  \centering
  \includegraphics[width=\textwidth]{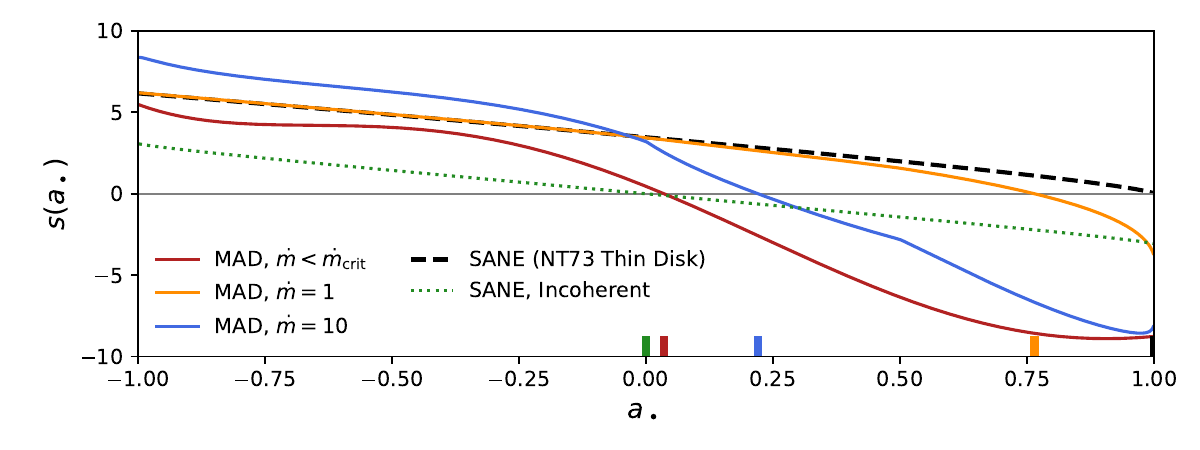}\\
  \caption{Spinup parameter ($s$; \autoref{eqn:s}) as a function of BH spin $a_\bullet$ for different types of accretion flows.  The \citet{Novikov&Thorne1973} thin disk, adopted for SANE accretion, is shown as a black dashed line.  MAD models with different values of $\dot{m}$ are shown with different colored solid lines.  As a dotted green line, we plot the SANE incoherent accretion case.  Equilibrium spins are marked on the x-axis for each type of accretion flow.}
  \label{fig:spinupExamples}
\end{figure*}

In \autoref{fig:spinupExamples}, we plot the spinup parameter $s$ as a function of BH spin $a_\bullet$ under different model assumptions.  MAD models with different values of $\dot{m}$ are plotted with different colored solid lines, revealing a strong dependence on both spin and $\dot{m}$. The equilibrium spin of each model is marked on the x-axis with the appropriate color.  Even at $\dot{m}=1$, the MAD model differs from the SANE one (dashed black line) in an important way, reaching an equilibrium spin of 0.75 rather than 1 for a thin disk.  We plot $s$ for incoherent SANE accretion as a dotted green line.  By construction, it is an odd function with an equilibrium spin of 0.  Interestingly, it can be seen that there are certain scenarios in which spin-down from the MAD model is more efficient than incoherent SANE accretion, such as $f_\mathrm{Edd}=10$ with $a_\bullet \gtrsim 0.3$.  

\subsection{Parameters Varied in This Study}

\begin{table*}[]
\begin{longtable}{|p{0.12\textwidth}|p{0.22\textwidth}|p{0.1\textwidth}|p{0.25\textwidth}|p{0.25\textwidth}|}
\hline
\textbf{Parameter}                   & \textbf{Astrophysical Question}                                                                          & \textbf{Values}               & \textbf{Effect on Mass Evolution}                                     & \textbf{Effect on Spin Evolution}                                                     \\ \hline
Magnetic Field State        & How magnetized are SMBH accretion flows?                                                          & \textbf{MAD}, SANE            & Indirect only, through accretion efficiency ($\epsilon$ and $\eta$).                & Modulates spin extraction through jet, dynamical effect on accretion flow.         \\ \hline
$p_\mathrm{merge}$          & What is the probability of SMBH merger after a major halo merger?                                 & 0, 0.1, \textbf{1}            & Directly scales the SMBH merger growth channel.             & SMBH mergers directly change the spin magnitude.                                   \\ \hline
Seeding                     & How are SMBHs initialized in the early universe?                                                & \textbf{Heavy}, Light         & Effect on assembly in dwarf halos and/or early universe. & Indirect only, since seeds are initialized at $a_\bullet=0$.                       \\ \hline
SMBH Merger Reorientability & Can major mergers cause retrograde accretion flows?                                                   & \textbf{Disruptive}, Gentle   & Indirect only, through accretion efficiency ($\epsilon$ and $\eta$).                & Disruptive mergers can produce retrograde systems.                                 \\ \hline
$\dot{m}_\mathrm{crit}$     & At what specific accretion rate do accretion flows transition from radiatively inefficient to efficient? & \textbf{0.003}, 0.03          & Indirect only, through accretion efficiency ($\epsilon$ and $\eta$).                & In MADs, strong magnetic fields produce more radial flows and BZ jets below this value. \\ \hline
Accretion Coherency         & Does the SMBH gas supply maintain a consistent angular momentum direction?                      & \textbf{Coherent}, Incoherent & Indirect only, through accretion efficiency ($\epsilon$ and $\eta$).                & Incoherent accretion implies 50\% of systems are retrograde, systematically spinning SMBHs down. \\ \hline
\end{longtable}
\caption{Summary of parameters explored in this study.  Throughout this work, we will refer to all variations of the model relative to the Fiducial model, whose values are bolded:  MAD $p_\mathrm{merge}=1$ Heavy Disruptive $\dot{m}_\mathrm{crit}=0.003$ Coherent.}
\label{tab:parameter_summary}
\end{table*}

In this work, we explore 6 free parameters, described in \autoref{tab:parameter_summary}.  The model we adopt as Fiducial is described by the following parameters:  MAD, with $p_\mathrm{merge}=1$, initialized with Heavy seeds, with Disruptive mergers, with $\dot{m}_\mathrm{crit}=0.03$, and with Coherent accretion, shown in boldface in the table.  Each parameter is chosen to explore one key astrophysical question of interest. Throughout this work, we will always refer to models relative to the Fiducial model; e.g., the SANE $p_\mathrm{merge}=0$ model has all of the values of the Fiducial model, except SANE and $p_\mathrm{merge}=0$, resulting in SANE $p_\mathrm{merge}=0$ Heavy Disruptive $\dot{m}_\mathrm{crit}=0.03$ Coherent.  Note that $p_\mathrm{merge}=0$ models are ruled out by the detection of the GW background, but they are useful for isolating the effects of BH-BH mergers in our models.

\begin{figure*}
  \centering
  \includegraphics[width=\textwidth]{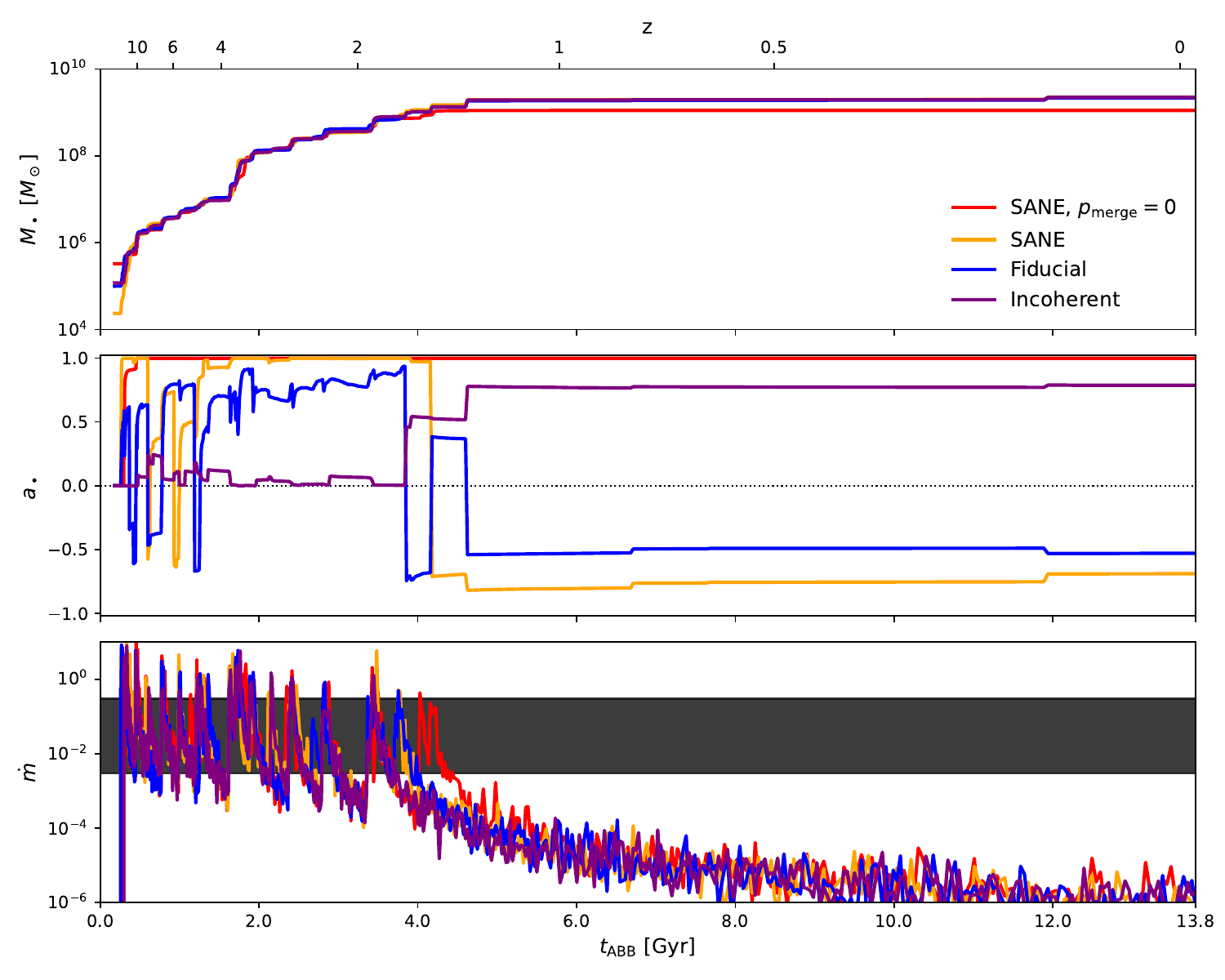}\\
  \caption{Example histories tracing back the most massive SMBHs in a $10^{14} \ M_\odot$ halo for a few model variants, SANE $p_\mathrm{merge}=0$, SANE, Fiducial, and Incoherent.  The remaining parameters are held fixed: heavy seeds, disruptive mergers, and $\dot{m}_\mathrm{crit}=0.003$.  In the black band ($0.3 \leq \dot{m} \leq \dot{m}_\mathrm{crit}$), even MAD models evolve as a thin disk, since there is insufficient magnetic flux to power an efficient jet \citep{Ricarte+2023d}.  In the model with incoherent accretion, only $a_\bullet > 0$ is plotted for clarity, but the sign of $a_\bullet$ is set randomly at each snapshot.}
  \label{fig:tracks}
\end{figure*}

Our fiducial model produces an acceptable $z=0$ $M_\bullet-\sigma$ relation and $0 \leq z \leq 6$ luminosity functions, shown in \autoref{sec:validation}. In \autoref{fig:tracks}, we plot example histories of BHs that become the most massive SMBHs in $10^{14} \ M_\odot$ halos at $z=0$, showing $M_\bullet$, $a_\bullet$, and $\dot{m}$ as a function of time for three different model variants.  Each dark matter merger tree is identical in this figure, leading to similar $\dot{m}$ histories.  We demarcate the approximate thin-disk regime for our MAD model in black, $\dot{m}_\mathrm{crit} < \dot{m} < 0.3$.  The SMBH in the SANE $p_\mathrm{merge}=0$ variant (red) rapidly spins up to 0.998 and remains at that value, since retrograde accretion is only ever triggered by BH mergers in this model, and in the absence of mergers in this instance the spin is unaltered.  The SANE model variant (orange) features these occasional disruptive merger-triggered sign flips in $a_\bullet$, but is usually able to recover and spin up to 0.998. The Fiducial model never reaches maximal spin, and is moderated by angular momentum losses from MAD jets.  This spin-down is apparently more effective than that reported by \citet{Massonneau+2023a}, who explored super-Eddington spin evolution in a hydrodynamical simulation of a high-redshift galaxy.  The difference likely lies in the details of the implementation and the accretion history.  In particular, using our equations derived from GRMHD, jet-mode feedback starts to become relevant not only at $\dot{m} > 1$, but already at $\dot{m} \gtrsim 0.3$.  Our results are qualitatively in line with \citet{Lupi+2024}, who applied the equations of \citet{Ricarte+2023d} in post-processing of an MBH in a high-redshift cosmological simulation.  Finally, the Incoherent model always leads to spin-down, acquiring non-zero spin only due to BH-BH mergers.

\section{Results}
\label{sec:results}

We present spin distributions for our Fiducial model and selected model variants.  We consider the different selection effects that may affect samples originating from X-ray reflection spectroscopy, ngEHT, and LISA. 

\begin{figure}[h]
  \centering
  \includegraphics[width=0.5\textwidth]{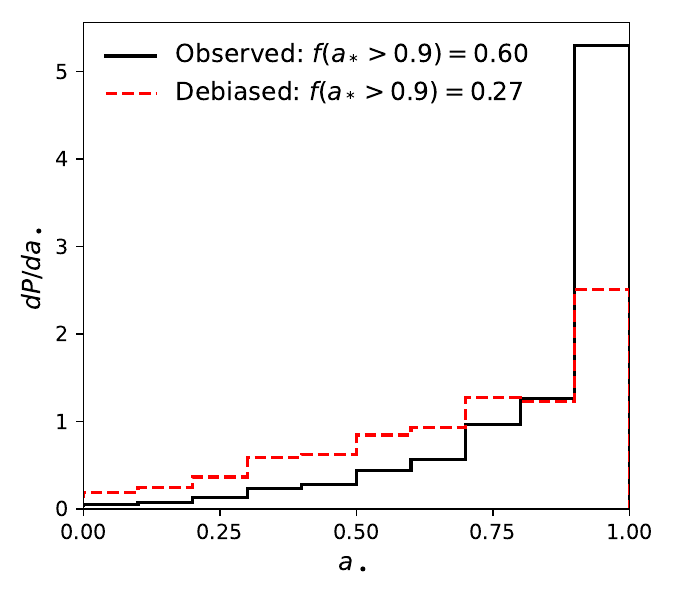}\\
  \caption{Spin distributions inferred from the observational data in \autoref{tab:spins_measured}.  The solid black curve bins the data with equal weights, while the dashed red curve attempts to remove a bias towards highly spinning sources due to spin-dependent radiative efficiency effects.  We write in the top left the fraction of objects with spin values above 0.9, which decreases after debiasing.}
  \label{fig:spins_observed}
\end{figure}

\subsection{High Accretion Rate Systems}
\label{sec:x-rays}

\begin{figure*}
  \centering
  \includegraphics[width=\textwidth]{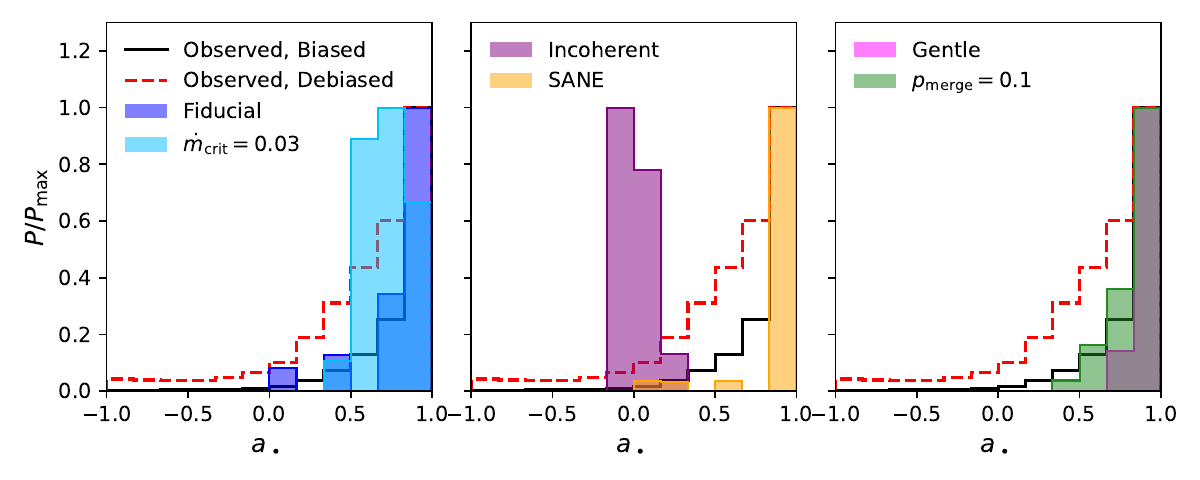}\\
  \caption{Comparison of spin distributions of model variants presented in \autoref{fig:spin_mass}, including only objects with $f_\mathrm{Edd} \in [0.01,1]$ at $z=0.1$.  Note that BHs are {\it not} weighted by cosmic abundance.  Our MAD Fiducial model variant successfully peaks at $a_\bullet \sim 1$ with some spread, but does not reproduce the width of the unbiased distribution.  The similarly performing $\dot{m}_\mathrm{crit}=0.3$ model variant more accurately reproduces the width of the unbiased distribution, but peaks at a more moderate spin of $a_\bullet \sim 0.7$.  Both the Incoherent and SANE variants are clearly inconsistent with the data.  The Gentle and $p_\mathrm{merge}=0.1$ variants show differences at low Eddington ratios in \autoref{fig:spin_mass}, but are indistinguishable at these higher Eddington ratios.}
  \label{fig:spin_Xref}
\end{figure*}

First, we consider SMBH spins that would be observable by X-ray reflection spectroscopy.  In \autoref{sec:spin_table}, we describe how we have compiled 55 spin measurements in the literature, producing both the observed and debiased distributions in \autoref{fig:spins_observed}.  This distribution peaks near $a_\bullet=1$, but has a broad tail of more moderate-spin objects.  We use these distributions for discussion and comparison, although as we discuss in \autoref{sec:spin_table}, these spin measurements bear significant model uncertainties and selection effects that discourage are more quantitative analysis.  In \autoref{fig:spin_Xref}, we compare {\sc Serotina} predictions with these distributions, including only objects with $f_\mathrm{Edd} \in [0.01,1]$.  This is an attempt to match selection effects, since the underlying assumption behind measuring spins from X-ray reflection spectroscopy is the existence of a radiatively efficient thin disk.  For this figure, in computing both the observed and theoretical distributions, we neglect a potential mass dependence.

\begin{figure*}
  \centering
  \includegraphics[width=\textwidth]{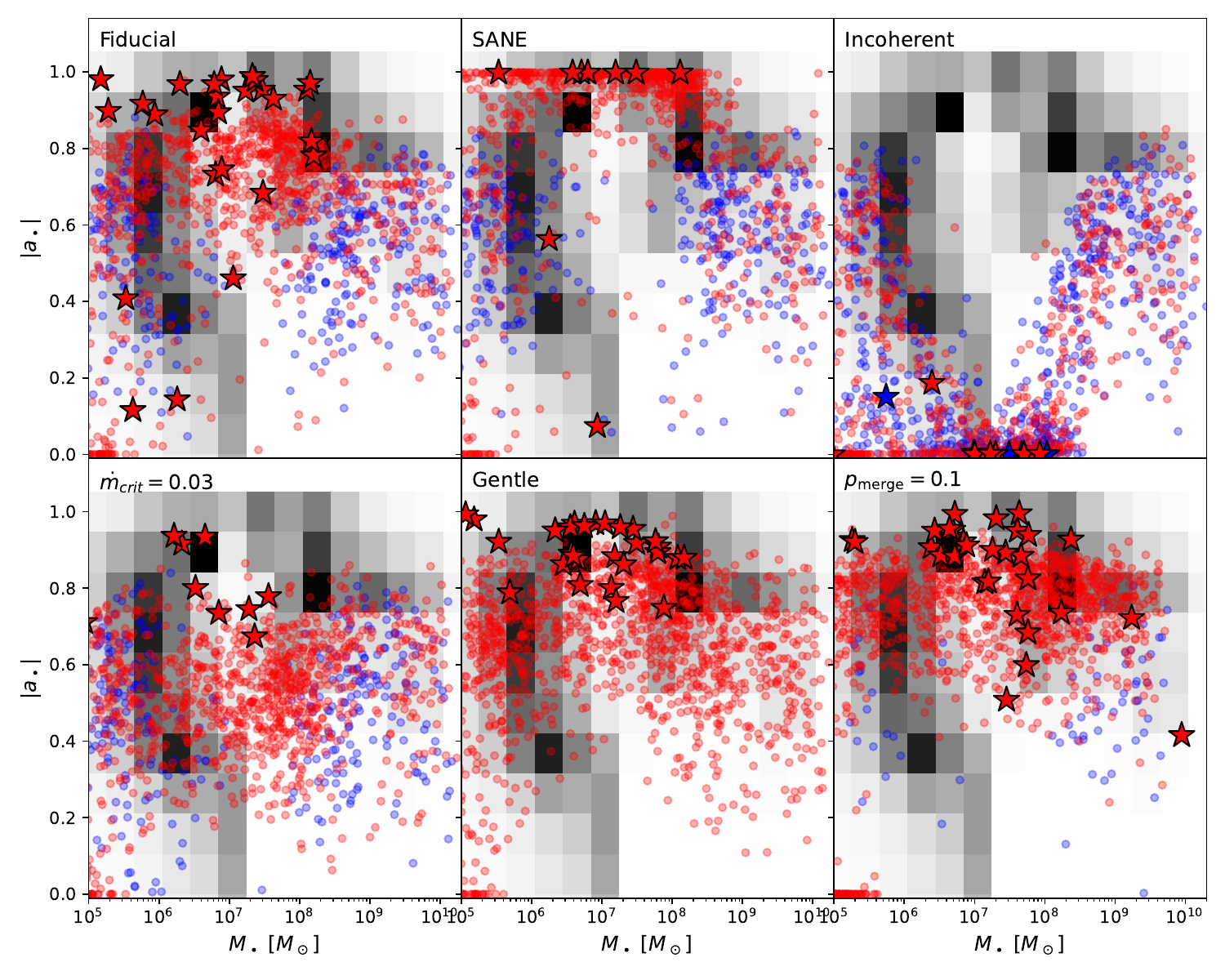}\\
  \caption{Comparison of spin as a function of mass in our model variants at $z=0.1$ (colored markers) with the distribution of claimed spin measurements from X-ray reflection spectroscopy compiled in \autoref{tab:spins_measured}.  Red markers are prograde, and blue markers are retrograde.  There have been no claims of a retrograde accretion disk observed by X-ray reflection spectroscopy.  To compare with the observed data, we mark objects with $f_\mathrm{Edd} \in [0.01,1]$ as stars.}
  \label{fig:spin_mass}
\end{figure*}

Then, in \autoref{fig:spin_mass}, we plot spin as a function of SMBH mass for our fiducial model and several model variants.  The X-ray reflection spectroscopy spin measurements compiled in \autoref{tab:spins_measured} have been binned into the greyscale distribution in the background.  The distribution is debiased with respect to radiative efficiency, as discussed in \autoref{sec:spin_table}.  Although $|a_\bullet|$ is plotted, \autoref{tab:spins_measured} does not contain any retrograde spin measurements.  Colored points represent {\sc Serotina} output, where a star is plotted instead of a circle if $f_\mathrm{Edd} \in [0.01,1]$.  Retrograde BHs are colored blue instead of red, revealing that although {\sc Serotina} can produce retrograde BHs, they usually do not stay retrograde if $f_\mathrm{Edd} > 0.01$.  Both plots represent $z=0.1$ rather than $z=0$ to provide better statistics.

Examining \autoref{fig:spin_Xref} and \autoref{fig:spin_mass}, our fiducial model is successful in qualitatively reproducing the observational trends that (1) there are no retrograde objects with $f_\mathrm{Edd} \in [0.01,1]$ and (2) there is a broad disribution that peaks near 1.  Comparing the Fiducial to the SANE model in \autoref{fig:spin_Xref}, we can see that BZ jets have a noticeable effect spinning down the SMBH population, since the SANE model BHs are more strongly clustered near $a_\bullet \approx 1$. This effect is noticeably stronger in the $\dot{m}=0.03$ model, which transitions to efficient jet spin-down at larger accretion rates.  In most models, even though there do exist retrograde systems at lower accretion rates (blue circles in \autoref{fig:spin_mass}), such systems are efficiently flipped to prograde if they accrete at high enough rates to be observed by X-ray reflection spectroscopy.  

The Incoherent model is clearly inconsistent with the distribution from X-ray reflection spectroscopy.  In this model, high Eddington ratio objects have $a_\bullet \approx 0$ and are equally likely to be prograde as they are retrograde.  We comment that although the equilibrium spin in our perfectly incoherent model is exactly $a_\bullet=0$, a different equilibrium can be reached depending on how much matter is assumed to come in each coherent chunk \citep[e.g.,][]{King+2008,Dotti+2013}.  In every model except the Incoherent accretion model, highly accreting objects are biased towards higher values of $|a_\bullet|$ for astrophysical reasons.

Next, we comment on how the spin distribution changes for finer details of the model.  The Gentle model demonstrates that disruptive mergers are solely responsible for the existence of retrograde BHs in the fiducial model in \autoref{fig:spin_mass}, but that these retrograde flips have little effect on the distribution of $|a_\bullet|$.  Finally, retrograde flips are less frequent in the $p_\mathrm{merge}=0.1$ variant, since they are tied to BH mergers. We also see that there are fewer BHs with $M_\bullet \sim 10^{10} \ M_\odot$ with the suppression of the merger growth channel.

\subsection{Low Accretion Rate Systems}
\label{sec:ngEHT}

The Event Horizon Telescope (EHT) collaboration has successfully resolved the polarized horizon-scale structure of two SMBHs \citep[e.g.,][]{EHTC+2019a,EHTC+2022a}.  By improving the coverage and sensitivity of the array on the ground (the next-generation EHT; ngEHT) and complementing the array with a station in space (Black Hole Explorer; BHEX), the array will gain access to dozens of additional targets \citep{Pesce+2021,Johnson+2024,Zhang+2024}.  Potential spin constraints are one of the science drivers of these developments \citep{Palumbo+2020,Emami+2023,Qiu+2023,Ricarte+2023b}.  The ground array aims to exploit a known link between resolved polarization structure, the magnetic field geometry, and the space-time to place model-dependent spin constraints on up to dozens of targets \citep{Palumbo+2020,Emami+2023,Qiu+2023}.  Spatially resolved dynamical motion will impose additional spin constraints on Sgr A$^*$ \citep{Conroy+2023}.  Finally, resolving the photon ring will provide an exquisite spin measurement of $\mathrm{M}\;87^*$ and possibly also Sgr A$^*$ with an expected 10\% measurement uncertainty, assuming only the Kerr metric \citep[e.g.,][]{Johannsen&Psaltis2010,Johnson+2020,Johnson+2024}.

Potential spin measurements from ngEHT and BHEX would have very different selection effects compared to X-ray Reflection Spectroscopy, selecting for nearby sources with $M_\bullet \gtrsim 10^8 \ M_\odot$ (for angular resolution) and {\it lower} Eddington ratios (for lower optical depth to see the shadow).  As discussed in \autoref{sec:x-rays} and shown in \autoref{fig:spin_mass}, we expect these two populations to have different spin distributions on physical grounds, with a lower Eddington ratio sample representing more typical SMBHs.

\begin{figure*}
  \centering
  \includegraphics[width=\textwidth]{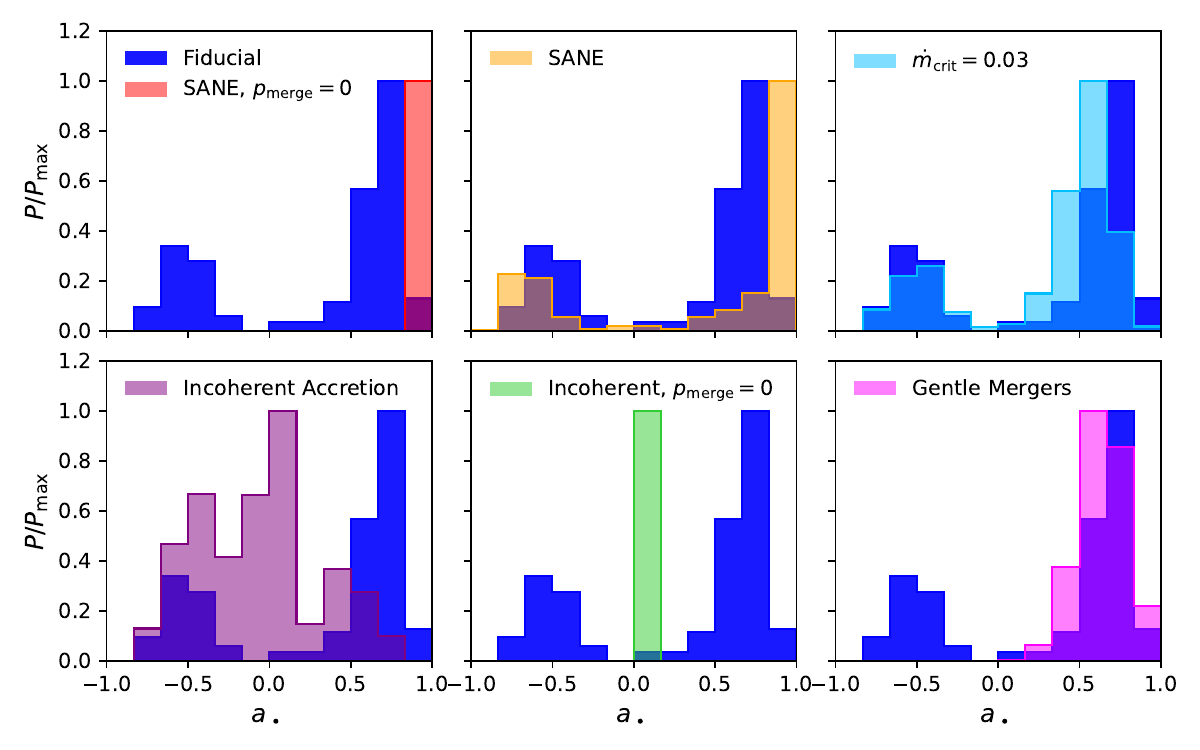}\\
  \caption{Spin distributions of SMBHs above $10^8 \ M_\odot$ with $\dot{m}<3\times 10^{-4}$ at $z=0.1$, mimicking the expected selection effects for observations with ngEHT and/or BHEX.  Similar astrophysics is encoded as in \autoref{fig:spin_Xref} and \autoref{fig:spin_mass}, but for a more representative population of BHs.
  \label{fig:ngEHT}}
\end{figure*}

In \autoref{fig:ngEHT}, we mimic these selection effects by computing spin distributions for objects with $M_\bullet > 10^8 \ M_\odot$ and $f_\mathrm{Edd} < 3 \times 10^{-4}$, based on the sample presented in \citet{Zhang+2024}.  Objects are weighted according to the number density of the host halo.  As in \autoref{fig:spin_mass}, we plot the $z=0.1$ snapshot for better statistics compared to the $z=0$ snapshot.  The MAD Fiducial model is shown in each panel as the blue histogram.  Even though this model reproduces the observations that all spins measured by X-ray reflection spectroscopy are prograde, this model contains some retrograde SMBHs due to the ``disruptive'' merger prescription.  This agrees with several cosmological simulations in the literature that find a larger fraction of retrograde systems at higher-masses \citep{Bustamante&Springel2019,Sala+2024}.  In the MAD Fiducial model, EHT would be unlikely to observe systems with $|a_\bullet| \lesssim 0.2$, and while it may observe retrograde systems, it would be unlikely to find systems with $a_\bullet < -0.9$.  

We compare 6 model variants against the MAD Fiducial model, showing substantial differences.
\begin{itemize}
    \item {\bf SANE $p_\mathrm{merge}=0$:}  This model has coherent accretion, no jet spin-down, and no BH mergers.  As a result, any BHs that have grown significantly from their seed masses have $a_\bullet=0.998$.
    \item {\bf SANE:}  Compared to the SANE $p_\mathrm{merge}=0$ variant, BH-BH mergers are sufficient to produce a wider distribution of spins at these high-masses (see also \autoref{fig:spin_mass}), in agreement with \citet{Sala+2024}.
    \item {\bf $\dot{m}_\mathrm{crit}=0.03$:}  These distributions are sensitive to the transition between thin disk and hot accretion, a currently poorly understood aspect of accretion disk theory.  Setting $\dot{m}_\mathrm{crit}=0.03$ instead of $\dot{m}_\mathrm{crit}=0.003$ shifts these distributions inwards, since AGN transition from thin disk spin-up into jet-driven spin-down earlier in their evolution.
    \item {\bf Incoherent Accretion:}  In this variant, spin-down happens at all Eddington ratios, causing the distribution to peak at $a_\bullet=0$.  While not shown, the equivalent SANE variant is similar.
    \item {\bf Incoherent Accretion, $p_\mathrm{merge}=0$:}  Turning off BH-BH mergers reveals that mergers are solely responsible for the width of the distribution in the previous variant.  Consequently, among Incoherent variants, the width of the distribution is sensitive to $p_\mathrm{merge}$.
    \item {\bf Gentle Mergers:}  In the MAD Fiducial model, disruptive mergers are the only mechanism that can change accretion flows from prograde to retrograde.  Thus, there are no retrograde SMBHs in this variant.
\end{itemize}

Given the dramatic differences between these variants, even a handful of robust spin measurements would be able to distinguish some of them.  For example, a single retrograde measurement would rule out the Gentle model, and a distribution of highly spinning BHs selected in this fashion would further disfavor the Incoherent model.

\subsection{Gravitational Waves}
\label{sec:lisa}

During the merger of two BHs, their spins couple with the orbital evolution, leading to a detectable signature imprinted in the gravitational waveform.  LISA will be able to exploit this signature, leading to BH spin measurements across cosmic time with sub-percent precision \citep{Berti+2005,Lang+2006,Lang+2007,Lang+2008,Lang+2011,Baibhav+2020,Bhagwat+2022,Watarai+2024}. LISA will also measure spins of SMBHs with sub-percent precision from extreme mass ratio inspirals (EMRIs), but we have excluded EMRIs from our current analysis since estimating the EMRI rate is not straightforward \citep{Babak+2017}.

\begin{figure*}
  \centering
  \includegraphics[width=\textwidth]{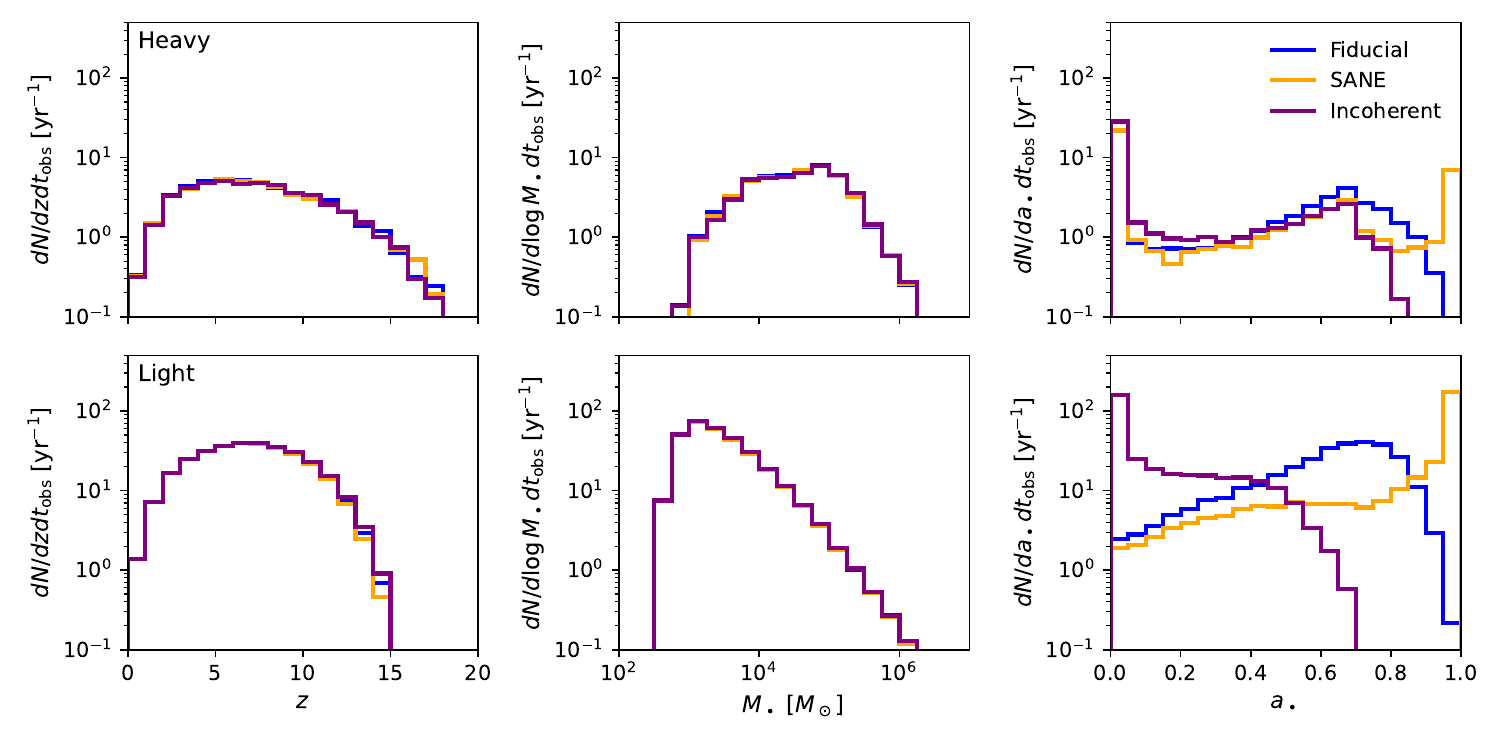}
  \caption{Distributions of LISA event rates for selected model variants.  The ability to distinguish seeding models via redshift and chirp mass distributions in {\sc Serotina} were highlighted in \citet{Ricarte&Natarajan2018b}. While the redshift and mass distributions are insensitive to a BH's accretion history, this is clearly imprinted on the spin distribution.  Models can especially be distinguished at the extremes of these distributions.}
  \label{fig:lisa}
\end{figure*}

In \autoref{fig:lisa}, we repeat the analysis of \citet{Ricarte&Natarajan2018b} and calculate the LISA event rate as a function of redshift, chirp mass, and now primary BH spin.  Compared to this previous work, we only included numbers for $p_\mathrm{merge}=1$ in light of the recently detected nHz gravitational wave background.\footnote{Note that in \citet{Ricarte&Natarajan2018b}, we labeled $p_\mathrm{merge}=1$ as ``optimistic.''}  Events are filtered using the LISA sensitivity curve such that only mergers that would be detected with a signal-to-noise ratio of 8 or larger are included.  We focus on 6 model variants.  In the top row, we plot variants with heavy seeds, and in the bottom row we plot variants with light seeds. The Fiducial model is shown in blue, with the SANE variant shown in orange and the Incoherent variant shown in purple.

The redshift and mass distributions of events are very similar to those presented in \citet{Ricarte&Natarajan2018b}. Mergers produced in this model may be observable even in the seeding epoch ($z \geq 15$). Light seeds produce more events overall due to their larger abundance, while heavy seeds produce events that can be seen at higher redshift.  This is because heavy seeds are for the most part born already in the LISA band, while light seeds need to accrete before becoming detectable. The light seed chirp mass function peaks at $\sim 10^3 \ M_\odot$, falling at lower masses only because of the LISA sensitivity curve, while the heavy seed chirp mass function peaks at around $\sim 10^5 \ M_\odot$ due to the initial mass function assumed \citep{Lodato&Natarajan2007}.  However, the redshift and mass distributions are completely insensitive to the manner in which mass has been accumulated.

The LISA spin distribution is critical for distinguishing variants, especially with respect to their accretion prescriptions. Two types of models produce a large population of $a_\bullet \approx 0$ mergers.  First, this is the most common spin for heavy seed models because we assume that the natal spin is 0, a natural consequence of a MAD super-Eddington formation mechanism. This signature is absent for most of the light seed models because they needed to accrete mass to reach the LISA band, erasing this signature.  In other words, the natal spin distribution may be imprinted on the spin distribution of LISA events preferentially for heavy seeds, but not for light seeds.  Any alternative initial spin distribution would also have appeared here \citep[see][for a similar effect]{Barausse2012}.  It is also possible to produce mostly $a_\bullet=0$ objects if accretion is incoherent, such that the equilibrium spin is always exactly 0, as seen in the Incoherent Light variant.  However, recall that such a scenario is disfavored by current X-ray reflection spectroscopy spin measurements (\autoref{sec:x-rays}) as seen in {\autoref{fig:spin_Xref}}.  

While the abundance of objects with $a_\bullet \sim 0$ most effectively distinguishes between light and heavy seeds, the abundance of objects with $a_\bullet \sim 1$ most effectively discriminates between MAD and SANE.  Our SANE model distributions always peak at $a_\bullet \approx 1$, since their accretion prescriptions push BHs towards $a_\bullet = 0.998$.  The MAD models peak at more moderate spins of $a_\bullet \sim 0.7$, due to spin-down both at $\dot{m} \gtrsim 0.3$ and $\dot{m} < \dot{m}_\mathrm{crit}$.  Naturally, the MAD Incoherent variants produce almost no mergers with $a_\bullet \gtrsim 0.8$.  While not shown, the SANE Incoherent variants are similar.

In summary, the LISA spin distributions of our model variants are easily distinguishable and bear signatures of our seeding and accretion prescriptions, especially at the extremes.  We also find that the natal spin distribution, assumed to be a delta function at 0 in our model, may be imprinted on the spin distribution of heavy seeds, but is likely to be erased for light seeds.  Event rates for each of our variants are tabulated in \autoref{tab:lisa}, where we include the event rates for only $a_\bullet < 0.2$ and $a_\bullet > 0.8$ separately.  Error bars are estimated through bootstrapping, and account only for cosmic variance, which is small since these estimates are integrated across the entire observable universe.  MAD variants have a smaller fraction of events at $a_\bullet > 0.8$ than their SANE counterparts.  Heavy variants have a much larger fraction of events with $a_\bullet < 0.2$ than their Light counterparts.  The Light Incoherent variant produces the most events with $a_\bullet < 0.2$ and the fewest with $a_\bullet > 0.8$, while the Light SANE Coherent variant produces the least events with $a_\bullet < 0.2$ and the most with $a_\bullet > 0.8$.  

\begin{table*}[]
\centering
\begin{tabular}{|lccc|}
\hline
Model Variant & Total Event Rate $[\mathrm{yr}^{-1}]$ & Event Rate $|a_\bullet| < 0.2$ $[\mathrm{yr}^{-1}]$ & Event Rate  $|a_\bullet| > 0.8$ $[\mathrm{yr}^{-1}]$\\
\hline
Heavy MAD Coherent   & $49.6 \pm 1.1$             & $24.1 \pm 0.8$          & $2.94 \pm 0.14$          \\
Heavy SANE Coherent  & $48.8 \pm 0.8$             & $24.0 \pm 0.6$          & $9.2 \pm 0.2$          \\
Heavy MAD Incoherent & $48.5 \pm 0.6$             & $32.0 \pm 0.6$          & $0.19 \pm 0.02$          \\
Light MAD Coherent   & $311 \pm 2$             & $13.8 \pm 0.6$          & $40.5 \pm 0.6$          \\
Light SANE Coherent  & $306 \pm 2$             & $10.0 \pm 0.4$          & $222 \pm 2$          \\
Light MAD Incoherent & $315 \pm 2$             & $218.2 \pm 1.3$          & $0.0 \pm 0.0$     \\
\hline
\end{tabular}
\caption{Observable LISA event rates for the model variants shown in \autoref{fig:lisa}.  Error bars are estimated via bootstrapping and only include cosmic variance.}
\label{tab:lisa}
\end{table*}

\section{Discussion}
\label{sec:discussion}

\subsection{Self-consistent Jet Power Predictions}
\label{sec:jet}

By combining equations \ref{eqn:eta_bz}, \ref{eqn:phi_hot}, and \ref{eqn:phi_superEdd}, our model self-consistently predicts the jet power of each SMBH without the addition of any free parameters.  According to our equations, efficient BZ jets occur at both $\dot{m} < \dot{m}_\mathrm{crit}$ and $\dot{m} \gtrsim 0.3$.  The jet power is not used in {\sc Serotina}, but could directly correspond to mechanical feedback in cosmological simulations or other models that track gas evolution.

\begin{figure}
    \centering
    \includegraphics[width=0.45\textwidth]{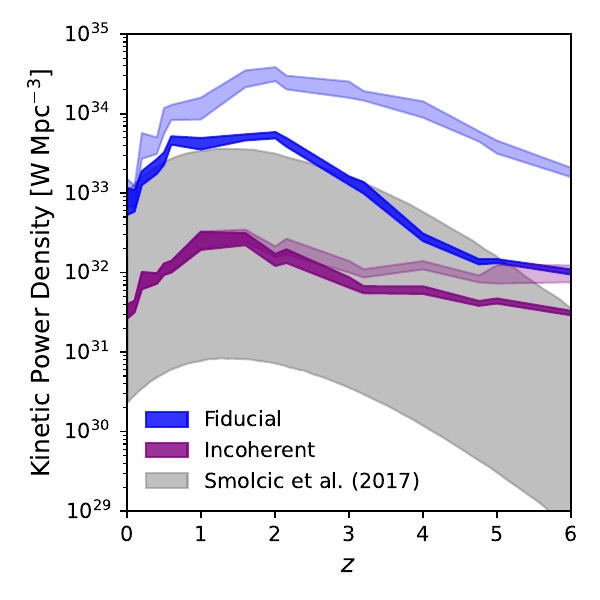}
    \caption{Integrated kinetic jet power density of our Fiducial and Incoherent models, computed using \autoref{eqn:eta_bz}.  The darker bands correspond to only objects with $\dot{m} < \dot{m}_\mathrm{crit}$, while the lighter bands also include super-Eddington jets.  We compare with loose, model-dependent constraints derived from radio luminosity functions in \citet{Smolcic+2017} in grey.  The conversion between radio luminosity and jet power is highly uncertain, and as discussed \citet{Smolcic+2017}, this broad grey band excludes the conversion proposed in \citet{Godfrey&Shabala2016} that would allow arbitrarily large values of the kinetic power density. Our Fiducial model pushes the upper limit of this band, while the Incoherent model lies in the middle.}
    \label{fig:kinetic_density}
\end{figure}

Jet power is difficult to access observationally, but we show that trustworthy estimates of jet power would allow us distinguish between MAD and SANE, or Coherent and Incoherent accretion flows.  As reviewed in \citet{Smolcic+2017}, many different relationships have been developed to translate from jet power to 1.4 GHz radio luminosity \citep{Willott+1999,Merloni&Heinz2007,Cavagnolo+2010,OSullivan+2011,Daly+2012,Heckman&Best2014,Godfrey&Shabala2016}.  These relationships vary by orders of magnitude.  Most concerningly, \citet{Godfrey&Shabala2016} found that Malmquist bias drove these relationships, and that only a weak correlation with large scatter remained after accounting for this bias.

We plot the integrated kinetic power density due to jets as a function of redshift for our Fiducial and Incoherent models in \autoref{fig:kinetic_density}.  By modeling the evolution of the radio luminosity function and exploring a variety of relationships for the conversion between 1.4 GHz radio luminosity and jet power, \citet{Smolcic+2017} compute the grey band.\footnote{Here, the minimum and maximum curves for their ``PLE'' model of their Appendix A is considered.}  As discussed in \citet{Smolcic+2017}, this band does not include the distance bias-corrected \citet{Godfrey&Shabala2016} relationship, which can predict arbitrarily high jet power densities.  

For our models, the darker region represents only systems with $\dot{m} \leq \dot{m}_\mathrm{crit}=0.003$, while the lighter region includes the jet power from all accretion rates.  This distinction is important for the Fiducial model, for which most of the jet power originates in super-Eddington systems, likely separate from samples of ``radio-mode'' AGN.  When counting only radiatively inefficient systems, the Fiducial model roughly agrees with the upper part of the band, corresponding to the conversion between radio luminosity of \citet{Merloni&Heinz2007}. Interestingly, the Incoherent model lies in the middle of this band, discounting the very poorly constrained behavior at $z \gtrsim 5$.  Recall that the Incoherent model produces spin distributions centered on 0 and is inconsistent with spin constraints from X-ray reflection spectroscopy.  Due to these lower spin values, the kinetic power predicted by this model is lower than the Fiducial model by over an order of magnitude at $z=0$.  In this model, super-Eddington jets contribute much less to the kinetic power density, since high-Eddington ratio systems are rapidly spun down in this model.  These large differences suggest that these two accretion models would have significantly different effective feedback efficiencies if implemented into a cosmological simulation.

Recall that our SANE models produce 0 jet power by construction, and are thus not included in \autoref{fig:kinetic_density}, but it is worth reflecting on their potential signatures.  In particular, although we have bracketed the extremes in magnetic flux, a model where only a minority of AGN are MAD could also lower the kinetic power densities predicted by our models.  Such a scenario has been proposed to explain differences in radio loudness in the AGN population \citep{Sikora&Begelman2013}.

\subsection{Mapping Parameters to Observables}
\label{sec:param_to_obs}

Each of the parameters varied in our study impact these observables in different ways, summarized here.
\begin{itemize}
    \item Magnetic Field State:  MAD models produce efficient BZ jets.  This moderates spin across cosmic time, reducing the population of $a_\bullet \approx 1$ objects in any dataset.    
    \item $p_\mathrm{merge}$:  In {\sc Serotina}, SMBH mergers dominate the low-redshift mass and spin evolution of the most massive SMBHs if $p_\mathrm{merge}=1$.  Increasing $p_\mathrm{merge}$ increases the spread in spin of the highest-mass SMBHs.
    \item Seeding:  We have shown that ``Light'' and ``Heavy'' seeds can be distinguished by LISA event rates.  Since heavy seeds are formed in the LISA band, their natal spin distribution may be imprinted on the distribution of LISA-detectable mergers if they successfully merge.
    \item SMBH Merger Reorientability:  By construction, models with ``disruptive'' mergers flip their accretion disks when a major BH-BH merger occurs in a retrograde orbit.  This allows us to form retrograde accretion disks, especially for high-mass SMBHs.
    \item $\dot{m}_\mathrm{crit}$:  Models with higher values of $\dot{m}_\mathrm{crit}$ become radiatively inefficient and larger specific accretion rates, by construction.  they also produce BZ jets and spin-down at higher accretion rates, causing their spins to be slightly lower.
    \item Accretion Coherency:  Models with incoherent accretion always spin down to 0 and can only acquire non-zero spin via BH-BH mergers.  Consequently, they have spin distributions that peak at 0, systematically lower radiative efficiencies, and weaker BZ jets if they are MAD.
\end{itemize}

\section{Conclusions}
\label{sec:conclusions}

We have implemented into the SAM {\sc Serotina} a recently developed set of formulae for MAD accretion flows, self-consistently predicting jet power and spin-down based on GRMHD simulations \citep{Ricarte+2023d}.  We have shown that spin-down from the \citet{Blandford&Znajek1977} mechanism has a significant impact on cosmological time scales, moderating SMBH spins across cosmic time at either very low and very high accretion rates.  The astrophysics included here allows us to make self-consistent predictions for observable spin distributions and jet powers. 

First, we assembled a SMBH spin distribution based on X-ray reflection spectroscopy measurements, correcting for a bias in the population due to the expected spin-dependent radiative efficiency.  If these measurements are accurate, we find that they clearly disfavor incoherent accretion models, which would spin SMBHs down too rapidly.  Instead, they are qualitatively consistent with our Fiducial model including coherent accretion and jet-driven spin-down.  Our SANE model produces spins more clustered near 1, while a model which transitions into the ADAF regime more rapidly ($\dot{m}_\mathrm{crit}=0.03$) produces fewer high-spin objects.

Second, we predict spin distributions accessible to the EHT and its extensions.  We demonstrate how efficiently accreting objects appropriate for X-ray reflection spectroscopy spin measurements are astrophysically biased towards more highly spinning prograde systems than the general population, even after correcting for a bias due to the spin-dependent radiative efficiency.  Spin distributions of EHT-accessible objects bear similar signatures of the underlying astrophysics, but are more representative of the population as a whole.

Finally, we determined the spin distributions of SMBHs accessible to LISA.  Although the distributions of mergers with respect to mass and redshift bear little information about how mass is accumulated, the distributions with respect to spin clearly distinguish between MAD vs. SANE, or coherent vs. incoherent accretion histories.  The strongest differences appear at the extremes of the distribution.  A peak near $a_\bullet=0$ occurs for either heavy seed or incoherent models, while a peak at $a_\bullet=1$ only occurs in SANE models.  Spin moderation due to BZ jets in MAD models can manifest as a peak at an intermediate value.

The accretion disk formulae described in \autoref{sec:methodology} can continue to be improved numerically. We have shown that our spin predictions are sensitive to the transition between thin disks and hot accretion, currently included only as an abrupt transition at $\dot{m}_\mathrm{crit} \in \{0.003,0.03\}$.  This transition should be modeled using GRMHD simulations.  In addition, we note that the radiative efficiency formulae adopted in this work are based mostly on analytic models which could also be updated with numerical work.  Interestingly, some MAD simulations in the super-Eddington regime predict unexpectedly {\it high} radiatively efficiencies for spinning SMBHs \citep{Curd&Narayan2023}.

Although {\sc Serotina} will continue to be improved in areas including seeding, dynamics, and the assignment of accretion rates, this work represents our best effort to model the inner accretion flow based on advances in resolved imaging and GRMHD simulations without tuning any parameters.  We have shown how SMBH spin distributions from X-ray reflection spectroscopy, resolved imaging, and LISA can be used to distinguish between different accretion histories.  In particular, spin-down due to the BZ mechanism in MADs should moderate spins in all of these samples.  This diversity of observational probes with distinct systematics and selection effects will be crucial to forming a robust and holistic picture of SMBH spin evolution.

\section{Acknowledgments}
This work was supported by the Black Hole Initiative at Harvard University, made possible through the support of grants from the Gordon and Betty Moore Foundation and the John Templeton Foundation. The opinions expressed in this publication are those of the author(s) and do not necessarily reflect the views of the Moore or Templeton Foundations.

\appendix

\section{The $M_\bullet-\sigma$ Relation and Bolometric Luminosity Functions}
\label{sec:validation}

\begin{figure*}
  \centering
  \includegraphics[width=\textwidth]{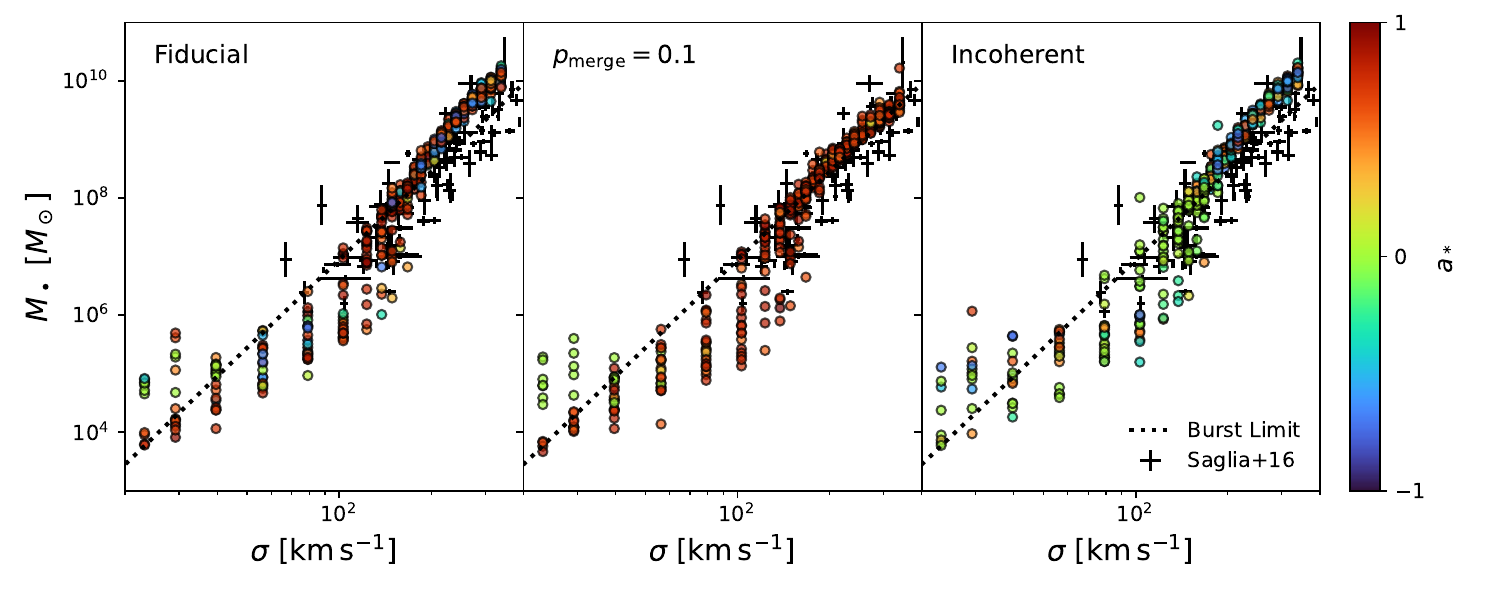}\\
  \caption{The $M_\bullet-\sigma$ relation at $z=0$ predicted by our fiducial model and a few variants, color-coded by $a_\bullet$. The burst growth limit of \autoref{eqn:M_max} is shown as a dotted line. Dynamically measured SMBH masses compiled by \citet{Saglia+2016} are shown as black errorbars. It is seen that the $M_\bullet-\sigma$ relation and its scatter for our Fiducial model is mostly spin-agnostic.}
  \label{fig:msigma}
\end{figure*}

\begin{figure*}
  \centering
  \includegraphics[width=\textwidth]{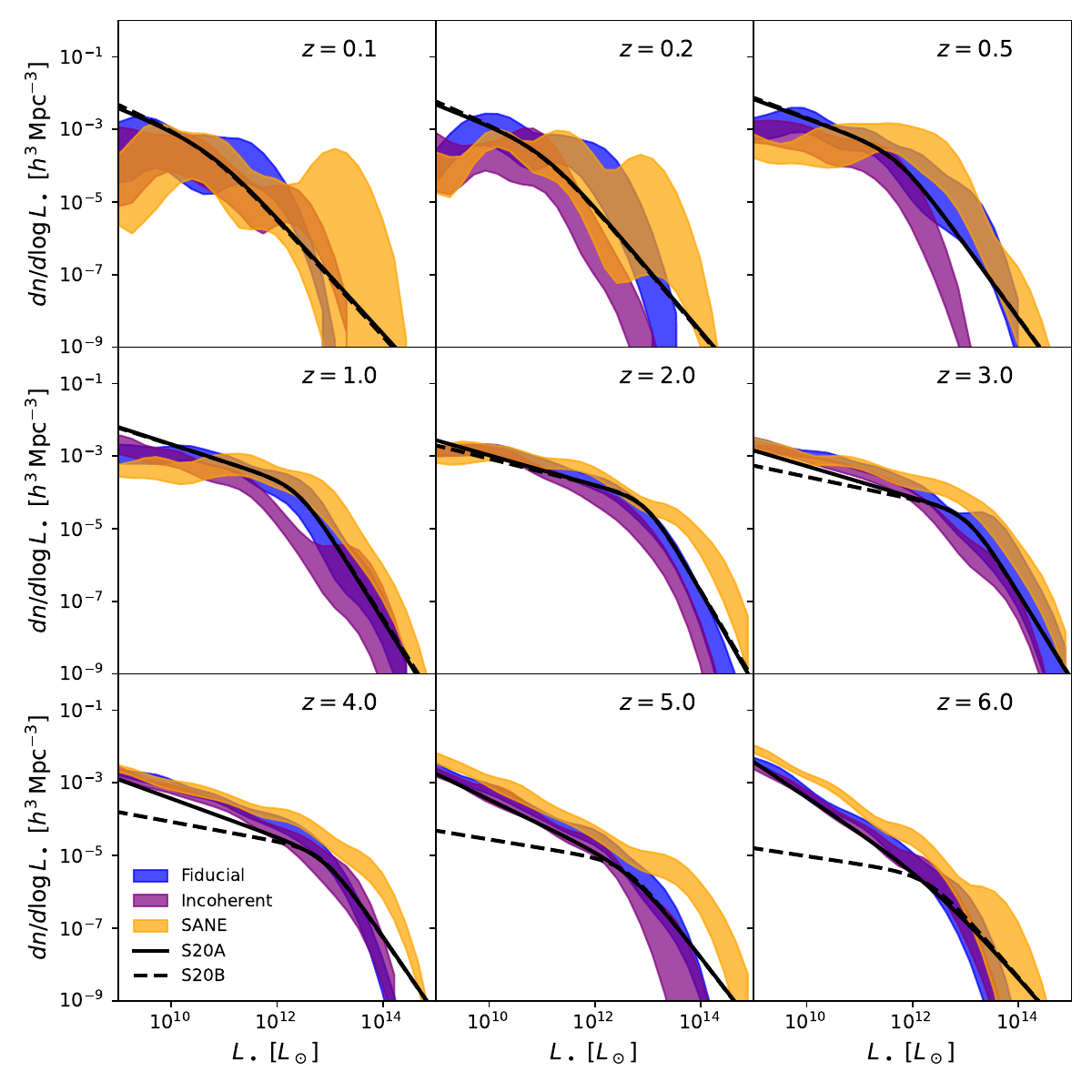}\\
  \caption{Bolometric luminosity functions of our fiducial model and a few variants compared to the observational data of \citet{Shen+2020}.  S20A corresponds to their ``global fit A,'' and S20B corresponds to their ``global fit B.''  Differences in spin cause differences in radiative efficiency that can cause slight differences in the luminosity function.}
  \label{fig:lum}
\end{figure*}

To assess the accuracy of our model, we plot the $M_\bullet-\sigma$ relation in \autoref{fig:msigma} and bolometric luminosity functions in \autoref{fig:lum}.  Recall that the two parameters in \autoref{eqn:M_max} are tuned to match these quantities.

The $M_\bullet-\sigma$ relation at $z=0$ from our fiducial model and a few variants are presented in \autoref{fig:msigma}.  Dynamical mass measurements compiled by \citet{Saglia+2016} are shown with black error bars.  The mass at which SMBHs transition from burst to decline mode (\autoref{eqn:M_max}) is shown as the dotted line.  Points are color-coded according to their spins.  The fiducial model broadly agrees with the observational data, but has additional structures.  There is more scatter at low-$\sigma$ than at high-$\sigma$ due to the relative infrequency of major mergers in lower-mass halos.  In some of the lowest-mass halos, BHs are over-massive with respect to \autoref{eqn:M_max} because they were seeded at masses greater than this value.  Indeed, many of them have retained their natal spin of 0.  At high-$\sigma$, SMBHs are systematically over-massive with respect to \autoref{eqn:M_max} and the observational data.  This motivated $p_\mathrm{merge}=0.1$ in \citet{Ricarte&Natarajan2018a}, which has noticeably better agreement in the central panel.  However, our fiducial model has $p_\mathrm{merge}=1.0$ due to the large amplitude of the nHz gravitational wave background.  The incoherent model in the rightmost column has much lower spins compared to the fiducial model, but this does not noticeably affect the resultant $M_\bullet-\sigma$ relation. 

We compare the bolometric luminosity functions from our fiducial model and a few variants with the observational \citet{Shen+2020} luminosity functions in \autoref{fig:lum}.  As in our previous work, these luminosity functions are convolved with a log-normal distribution with width 0.3 dex in post-processing, to reproduce the intrinsic scatter of the $M_\bullet-\sigma$ relation, which strongly affects the behavior of the high-luminosity end.  Because the faint-end at high-redshift is poorly sampled and therefore sensitive to how this region is extrapolated, \citet{Shen+2020} provide two different fits: fit ``A'' where the faint end slope is freely allowed to evolve, and fit ``B'' where the faint end slope is forced to evolve monotonically with redshift.  All of our models achieve better agreement with fit ``A,'' especially at high-redshift.  All of the models struggle more at lower redshifts, where we are limited by poorer statistics as our merger trees contain more progenitors as redshift increases.  Since we fix $\dot{m}$, the dependence of the radiative efficiency on $a_\bullet$ leads to noticeable differences between the Fiducial, Incoherent, and SANE models at the high-luminosity end.  The SANE model SMBHs have spins up to $a_\bullet=0.998$, resulting in a greater abundance of high-luminosity AGN.  Meanwhile, the Incoherent model has systematically lower spins, resulting in systematically fainter AGN.

\section{Spin Measurements from X-ray Reflection Spectroscopy}
\label{sec:spin_table}

Although spins inferred from X-ray reflection spectroscopy are biased towards luminous sources and bear significant model uncertainties, they are nevertheless the most direct spin measurements available for AGN for theoretical comparison \citep[see][for recent reviews]{Reynolds2021,Bambi+2021}.  These measurements are made by fitting X-ray spectra using models that contain a spin-dependent reflecting surface, based on thin accretion disk theory.  We compile an inclusive list of 55 SMBH spin measurements in the literature in \autoref{tab:spins_measured}.  This is constructed beginning with recent compilations by \citet{Reynolds2021} and \citet{Bambi+2021}, to which a few more recent publications are added.  Note that \citet{Reynolds2021} and \citet{Bambi+2021} are reviews that include references to the original measurements.

\begin{table}[!ht]
	\centering
        \resizebox{!}{4in}{
	\begin{tabular}{|llll|}
		\hline
		Object & $M_\bullet \ [M_\odot]$ & $a_\bullet$ & Reference \\
		\hline
		Mrk 359 & $1.1 \times 10^{6}$ & $0.66^{+0.30}_{-0.54}$ & \cite{Bambi+2021} \\
		Ark 564 & $1.1 \times 10^{6}$ & >0.9 & \cite{Bambi+2021} \\
		Mrk 766 & $3.50^{+0.40}_{-0.30} \times 10^{6}$ & >0.92 & \cite{Bambi+2021} \\
		NGC 4051 & $1.91^{+0.78}_{-0.78} \times 10^{6}$ & >0.99 & \cite{Bambi+2021} \\
		NGC 1365 & $6.5^{+8.0}_{-3.5} \times 10^{7}$ & >0.97 & \cite{Bambi+2021} \\
		1H0707-495 & $6.5^{+8.0}_{-6.0} \times 10^{5}$ & >0.988 & \cite{Bambi+2021} \\
		MGC-6-30-15 & $2.9^{+1.8}_{-1.6} \times 10^{6}$ & $0.91^{0.06}_{-0.07}$ & \cite{Reynolds2021} \\
		NGC 5506 & $5.1^{+2.2}_{-1.2} \times 10^{6}$ & $0.93^{0.04}_{-0.04}$ & \cite{Bambi+2021} \\
		IRAS13224-3809 & $7.5^{+2.0}_{-2.5} \times 10^{5}$ & >0.975 & \cite{Bambi+2021} \\
		Ton S180 & $9.5^{+3.6}_{-2.9} \times 10^{6}$ & <0.37 & \cite{Bambi+2021} \\
		ESO 362-G18 & $1.25^{+0.45}_{-0.45} \times 10^{7}$ & >0.92 & \cite{Bambi+2021} \\
		Swift J2127.4+5654 & $1.5 \times 10^{7}$ & $0.72^{0.14}_{-0.20}$ & \cite{Reynolds2021} \\
		Mrk 335 & $1.78^{+0.46}_{-0.37} \times 10^{7}$ & >0.99 & \cite{Reynolds2021} \\
		Mrk 110 & $2.51^{+0.61}_{-0.61} \times 10^{7}$ & >0.99 & \cite{Bambi+2021} \\
		NGC 3783 & $2.98^{+0.54}_{-0.54} \times 10^{7}$ & >0.88 & \cite{Bambi+2021} \\
		1H0323+342 & $3.4^{+0.9}_{-0.6} \times 10^{7}$ & >0.9 & \cite{Bambi+2021} \\
		NGC 4151 & $4.57^{+0.57}_{-0.47} \times 10^{7}$ & >0.9 & \cite{Reynolds2021} \\
		Mrk 79 & $5.24^{+1.44}_{-1.44} \times 10^{7}$ & >0.5 & \cite{Bambi+2021} \\
		PG1229+204 & $5.7^{+2.5}_{-2.5} \times 10^{7}$ & $0.93^{0.06}_{-0.02}$ & \cite{Bambi+2021} \\
		IRAS13197-1627 & $6.4 \times 10^{7}$ & >0.7 & \cite{Bambi+2021} \\
		3C 120 & $6.9^{+3.1}_{-2.4} \times 10^{7}$ & >0.95 & \cite{Bambi+2021} \\
		Mrk 841 & $7.9 \times 10^{7}$ & >0.52 & \cite{Bambi+2021} \\
		IRAS09149-6206 & $1^{+3.0}_{-0.7} \times 10^{8}$ & $0.94^{0.02}_{-0.07}$ & \cite{Bambi+2021} \\
		Ark 120 & $1.5^{+0.19}_{-0.19} \times 10^{8}$ & $0.83^{0.05}_{-0.03}$ & \cite{Bambi+2021} \\
		RBS1124 & $1.8 \times 10^{8}$ & >0.97 & \cite{Bambi+2021} \\
		RXS J1131-1231 & $2 \times 10^{8}$ & $0.87^{0.08}_{-0.15}$ & \cite{Bambi+2021} \\
		Fairall 9 & $2.55^{+0.56}_{-0.56} \times 10^{8}$ & >0.997 & \cite{Bambi+2021} \\
		1H0419-577 & $1.30 \times 10^{8}$ & $0.96^{0.02}_{-0.00}$ & \cite{Bambi+2021} \\
		PG0804+761 & $5.5^{+0.6}_{-0.6} \times 10^{8}$ & >0.94 & \cite{Bambi+2021} \\
		Q2237+305 & $1 \times 10^{9}$ & $0.74^{0.06}_{-0.03}$ & \cite{Bambi+2021} \\
		PG2112+059 & $1 \times 10^{9}$ & >0.86 & \cite{Bambi+2021} \\
		H1821+643 & $4.5^{+1.5}_{-1.5} \times 10^{9}$ & >0.4 & \cite{Bambi+2021} \\
		IRAS 00521-7054 & $5 \times 10^{7}$ & >0.77 & \cite{Bambi+2021} \\
		IRAS 13349+2438 & $1^{+0.2}_{-0.4} \times 10^{7}$ & $0.3^{0.2}_{-0.5}$ & \cite{Bambi+2021} \\
		Fairall 51 & $1 \times 10^{8}$ & $0.8^{0.2}_{-0.2}$ & \cite{Bambi+2021} \\
		Mrk 1501 & $1.84^{+0.27}_{-0.27} \times 10^{8}$ & >0.98 & \cite{Bambi+2021} \\
		4C74.26 & $4^{+7.5}_{-2.5} \times 10^{9}$ & >0.5 & \cite{Bambi+2021} \\
		HE 1136-2304 & $7.9^{+4.5}_{-2.5} \times 10^{6}$ & >0.995 & \cite{Bambi+2021} \\
		Mrk 509 & $1.43^{+0.12}_{-0.12} \times 10^{8}$ & >0.993 & \cite{Bambi+2021} \\
		Mrk 1044 & $2.8^{+0.9}_{-0.7} \times 10^{6}$ & $0.997^{0.016}_{-0.001}$ & \cite{Bambi+2021} \\
		ESO 033-G002 & $3.2^{+3.1}_{-1.6} \times 10^{7}$ & >0.96 & \cite{Walton+2021} \\
		H1821+643 & $3.0^{+3.0}_{-1.8} \times 10^{9}$ & $0.62^{0.22}_{-0.37}$ & \cite{Sisk-Reynes+2022} \\
		J0107 & $1.6^{+1.6}_{-0.8} \times 10^{6}$ & $0.87^{0.08}_{-0.24}$ & \cite{Mallick+2022} \\
		J0228 & $3.2^{+3.1}_{-1.6} \times 10^{5}$ & $0.82^{0.16}_{-0.09}$ & \cite{Mallick+2022} \\
		J0940 & $1.6^{+1.6}_{-0.8} \times 10^{6}$ & $0.996^{0.001}_{-0.015}$ & \cite{Mallick+2022} \\
		J1023 & $5.0^{+5.0}_{-2.5} \times 10^{5}$ & $0.53^{0.39}_{-0.15}$ & \cite{Mallick+2022} \\
		J1140 & $1.26^{+1.25}_{-0.63} \times 10^{6}$ & $0.975^{0.012}_{-0.016}$ & \cite{Mallick+2022} \\
		J1347 & $1.0^{+1.0}_{-0.5} \times 10^{6}$ & $0.77^{0.19}_{-0.43}$ & \cite{Mallick+2022} \\
		J1357 & $1.6^{+1.6}_{-0.8} \times 10^{6}$ & $0.35^{0.15}_{-0.09}$ & \cite{Mallick+2022} \\
		J1434 & $6.3^{+6.3}_{-3.1} \times 10^{5}$ & $0.63^{0.27}_{-0.45}$ & \cite{Mallick+2022} \\
		J1541 & $1.6^{+1.6}_{-0.8} \times 10^{6}$ & $0.91^{0.07}_{-0.21}$ & \cite{Mallick+2022} \\
		J1626 & $5.0^{+5.0}_{-2.5} \times 10^{5}$ & $0.68^{0.28}_{-0.21}$ & \cite{Mallick+2022} \\
		J1631 & $6.3^{+6.3}_{-3.1} \times 10^{5}$ & $0.74^{0.16}_{-0.19}$ & \cite{Mallick+2022} \\
		J1559 & $1.6^{+1.6}_{-0.8} \times 10^{6}$ & $0.998^{+0.000}_{-0.023}$ & \cite{Mallick+2022} \\
		POX 52 & $3.2^{+1.0}_{-1.0} \times 10^{6}$ & $0.56^{0.36}_{-0.46}$ & \cite{Mallick+2022} \\
		\hline
	\end{tabular}
        }
	\caption{SMBH spin measurements compiled from the literature using X-ray reflection spectroscopy.  \citet{Bambi+2021} and \citet{Reynolds2021} correspond to compilations which contain the original references from which these values have been taken.}
	\label{tab:spins_measured}
\end{table}

Using the data from \autoref{tab:spins_measured}, we produce a spin distribution by superposing normalized probability distributions of each measurement, weighted equally. If a measurement is presented with error bars, then we assume that its probability distribution can be described by a Gaussian.  If the error bars are asymmetric, then a different Gaussian width is adopted below and above the measured value, and the distribution is renormalized.  For upper and lower limits, we assume uniform probability distributions either above or below the limit appropriately.  The resulting observed distribution is integrated into 10 equally spaced bins, and is presented as the black histogram in \autoref{fig:spins_observed}.  It has been often noted that observed spin measurements clearly prefer values nearer to 1 \citep[e.g.,][]{Reynolds2021,Bambi+2021}.  

Due to the exquisite X-ray spectra required, these objects are necessarily biased towards more luminous objects, which are expected to be the most highly spinning objects due to their larger radiative efficiencies \citep{Brenneman+2011,Vasudevan+2016}.  This results in a type of Malmquist bias.  In a flux limited survey, the number of objects of a given luminosity $N(L)$ in the sample is biased, per $N(L) \propto L^{3/2}$.  The distance to which an object of luminosity $L$ can be observed scales as $L^{1/2}$, and therefore the volume out to which such objects are accessible scales as $L^{3/2}$.  Since $L \propto \epsilon(a_\bullet)$, we obtain $N(L) \propto \epsilon(a_\bullet)^{3/2}$.  In the red dashed line of \autoref{fig:spins_observed}, we attempt to remove this spin-dependent bias by dividing the observed distribution by $\epsilon(a_\bullet)^{3/2}$, where $\epsilon(a_\bullet)$ for a thin disk is obtained via \citet{Novikov&Thorne1973}.  In the observed distribution, the fraction of objects with $a_\bullet>0.9$ is 0.60, but this drops to 0.27 in the de-biased distribution, which is much more uniform.  In \autoref{fig:spin_mass}, we perform a similar procedure in the two dimensions of mass and spin.  If a mass measurement is provided without error bars, then we conservatively assume standard deviations of 1 dex.  When comparing with {\sc Serotina}, we seek model variants that produce distributions peaking near maximal spin, but which also exhibit a substantial tail of sub-extremal BHs.

\bibliography{ms}

\end{document}